\documentclass[footinclude=false,11pt]{scrartcl}

\usepackage{fullpage}
\usepackage{amsmath}
\usepackage{amssymb}
\usepackage{xcolor}
\usepackage{graphicx}
\usepackage{url}
\RequirePackage[tt=false]{libertine}
\RequirePackage[varqu]{zi4}
\RequirePackage[libertine]{newtxmath}
\RequirePackage[T1]{fontenc}

\definecolor{myred}{rgb}{0.86,0.00,0.00}
\definecolor{myredlight}{rgb}{0.97,0.75,0.75}
\definecolor{myredlighter}{rgb}{0.99,0.94,0.94}
\definecolor{myredlighterr}{rgb}{1.0,0.98,0.98}
\definecolor{myblue}{rgb}{0.00,0.20,0.70}
\definecolor{mybluelight}{rgb}{0.75,0.80,0.93}
\definecolor{mybluelighter}{rgb}{0.94,0.95,0.98}
\definecolor{mybluelighterr}{rgb}{0.98,0.99,1.0}
\definecolor{mygreen}{rgb}{0.10,0.50,0.10}
\definecolor{mygreenlight}{rgb}{0.78,0.88,0.78}
\definecolor{mygreenlighter}{rgb}{0.94,0.97,0.94}
\definecolor{mygreenlighterr}{rgb}{0.99,0.99,0.99}
\definecolor{mygrey}{rgb}{0.40,0.40,0.40}
\definecolor{mygreylight}{rgb}{0.85,0.85,0.85}
\definecolor{mygreylighter}{rgb}{0.96,0.96,0.96}
\definecolor{mygreylighterr}{rgb}{0.99,0.99,0.99}
\definecolor{myorange}{rgb}{1.0,0.50,0.00}
\definecolor{myorangelight}{rgb}{1.0,0.87,0.75}
\definecolor{myorangelighter}{rgb}{1.0,0.96,0.93}
\definecolor{myorangelighterr}{rgb}{1.0,0.99,0.98}
\usepackage{algorithm}
\usepackage{algpseudocode}

\usepackage[nosfdefault]{comicneue} %PAF!
\usepackage{tikz}
\usepackage{pgfplots}
\usepgfplotslibrary{patchplots}
\usetikzlibrary{shadows,backgrounds,arrows,spy,decorations.pathreplacing}

\newcommand{\bh}{\mathbf{h}}
\newcommand{\bn}{\mathbf{n}}

\newcommand{\dd}{\text{d}}

\newcommand{\w}{\boldsymbol{\omega}}

\newcommand{\ind}{\mathbf{1}}
\newcommand{\Ppois}{\mathcal{P}}

\begin{document}

\title{An Elementary Expression for Multiple Scattering in Homogeneous Microflake Media}
\author{Jonathan Dupuy}
\date{August 2026}
\maketitle

\begin{abstract}
We derive a novel diffuse-like BRDF based on microfacet theory. Its main novelty lies in the fact that it 
admits an elementary expression for both evaluation and importance sampling, while accounting 
exactly for all scattering orders under the Smith shadowing assumption. 
As such, it is the first to settle the question of whether such a construction was possible.
In practice, the BRDF produces Lambertian reflectance at normal incidence and progressively concentrates 
reflected light around the mirror direction as incidence approaches grazing. We showcase this effect by comparing 
rendering results against Lambertian and state-of-the-art stochastic microfacet BRDFs.
\end{abstract}

\section{Introduction}

\paragraph*{Context}
Microfacet theory is a geometric optics model linking surface reflectance to
specific arrangements of microscopic facets that compose the surface and deviate incident 
light\footnote{This idea is quite old as it dates back to the work of 
Bouguer~\cite{bouguer1760} according to Trowbridge and Reitz~\cite{trowbridge1975}.}. 
This theory was introduced to the computer 
graphics community by Blinn~\cite{blinn1977reflection} and has undergone considerable development 
ever since. In this work, we contribute to this development by deriving a BRDF~\cite{nicodemus1992} from the 
multiple reflections among a microfacet arrangement subject to the state-of-the-art Smith shadowing 
assumption~\cite{smi_1967,Heitz2014Microfacet}.

\paragraph*{Smith Microfacets}
The Smith shadowing assumption treats microfacet orientations as statistically independent of their positions over the surface~\cite{bro_1983}.
In practice, this makes microfacet transport equivalent to that of a participating medium, 
with the BRDF describing the directional distribution of rays escaping the medium's boundary~\cite{heitz2016multiple,dupuy2016unification}. 
Unfortunately, this directional distribution is hard to derive for the complete transport~\cite{bitterli2022positionfree,cui2023invariance}, 
and so microfacet BRDFs either restrict transport to single-scattering effects or rely on a stochastic representation.
This dichotomy creates a trade-off between analytic simplicity and physical completeness: 
single-scattering models admit elementary closed-form expressions for evaluation and sampling but lose energy by neglecting inter-reflections, 
whereas multiple-scattering models preserve energy but lack an elementary closed-form characterization. 
Whether a Smith microfacet BRDF can combine exact multiple scattering with an elementary closed-form expression 
has remained an open question.

\paragraph*{Contribution and Outline}
In this work, we derive a particular Smith microfacet BRDF that answers this question in the affirmative:
\begin{equation}
\boxed{
f_r(\w_i,\w_o)=\frac{z_i+z_o}{\pi\left(1+\w_i\cdot\w_o\right)}.
}
\label{eq:main-brdf}
\end{equation}
This BRDF accounts exactly for all scattering orders and preserves energy, while retaining the elementary evaluation 
and direct importance sampling usually associated with single-scattering microfacet models. 
In the remaining sections, we derive this result in a self-contained way as follows:
\begin{itemize}
\item In Section~\ref{sec:generic}, we introduce the homogeneous semi-infinite microflake model and show how the normal distribution function 
(NDF) fully specifies its stochastic transport.
We then introduce a one-dimensional depth parameterization from which the escape probabilities associated with the first two scattering orders follow directly.
%We then derive the first two scattering orders and recover those of the classical Chandrasekhar BRDF.
\item In Section~\ref{sec:heightfield}, we show that one-sided NDFs reproduce the characteristic transport of mirror heightfields.
This structure allows us to derive the escape probability associated with any prescribed sequence of directions.
By marginalizing this probability over the intermediate directions, we obtain a scattering-order representation of the exit-direction distribution.
%We instantiate the first two orders for uniform-cap scattering, corresponding to unit-roughness GGX, and show why this simple choice does not lead to an elementary all-orders expression.
\item In Section~\ref{sec:quadratic}, we introduce a parameter-free quadratic NDF for which the intermediate path directions can be marginalized analytically and the complete scattering-order series can be summed in closed form, leading to Equation~\eqref{eq:main-brdf}.
\item In Section~\ref{sec:sampling}, we interpret the resulting BRDF as the Jacobian of a chordal disk transformation and use this construction to sample both the complete BRDF 
and its phase function.
\item In Section~\ref{sec:results}, we compare rendering results of our BRDF against Lambertian, single-scattering, and stochastic microfacet BRDFs.
\end{itemize}

% =============================================================================================
% =============================================================================================
% =============================================================================================
\section{Homogeneous Semi-Infinite Microflake Media}
\label{sec:generic}

In this section we introduce the homogeneous semi-infinite microflake medium as a starting primitive for 
modelling reflectance. 

% =============================================================================================
% =============================================================================================
\subsection{Preliminaries}
\label{sec:generic:preliminaries}

% =============================================================================================
\paragraph{Random Walks in Semi-Infinite Media}
A surface BRDF associates each incident direction with a distribution of reflected energy over outgoing directions.
An energy-preserving BRDF can therefore be interpreted as a stochastic mapping from a direction pointing \emph{towards} the surface to a random direction pointing \emph{away} from it.
A random walk in a semi-infinite medium naturally implements such a mapping because any ray crossing the medium boundary necessarily does so along an upwelling direction.
Algorithm~\ref{alg_walk} implements this construction and is illustrated in Figure~\ref{fig_geometry}.
Given a surface with normal $\bn := (x_n,y_n,z_n)\in\mathcal S^2$ and a ray with downwelling direction $\w_1$, i.e., 
$(\w_1\cdot\bn)\leq0$, the algorithm generates a random walk composed of $k\geq2$ free-flight distances $\tau_k=(t_1,\ldots,t_{k})$ and directions 
$\Gamma_k=(\w_1,\ldots,\w_{k})$.
Such a walk thus contains $k-1$ collisions, with the last free flight taking the ray outside the medium.
The walk terminates when the ray exits the medium and returns the final upwelling direction $\w_o:=\w_{k}$, with $(\w_o\cdot\bn)>0$.

% =============================================================================================
\begin{figure}[htbp]
    \centering
    \begin{minipage}[t]{0.45\linewidth}
        \vspace{0pt}
        \centering
        \resizebox{\linewidth}{!}{\pgfplotsset{every axis/.append style = {
    x = 1.75cm,
    y = 1.75cm,
    xmin = -2, xmax = 2,
    ymin = -1.5, ymax = 1,
    grid = both,
    axis line style = {color = black},
    minor grid style={dotted},
    major grid style={dotted},
    xticklabels = {},
    yticklabels = {},
    y label style = {rotate = -90},
    minor tick num = 3,
    ytick = {1,0,-1}
}}
\centering
\begin{tikzpicture}
\newcommand{\raysegment}[2]{%
    % segment fléché
    \draw[myred, ->, >=stealth] (#1) -- (#2);
    
    % accolade raccourcie
    \draw[myred, decorate,
          decoration={brace, raise=3pt,
                      pre=moveto, post=moveto,
                      pre length=4pt, post length=3pt}]
    (#2) -- (#1);
}

\pgfmathsetseed{12346}
\begin{axis}[
    no markers,
    name = myaxis
]
% macroscopic medium
\addplot+[fill, color = myblue, opacity = 0.2] coordinates
{(-3,0) (3,0) (3,-3) (-3,-3)} --cycle;
\addplot [myblue] {0};
\node [draw = none, myblue, anchor = north east] at (axis cs:-0.75,0.20) {\scalebox{0.65}{medium interface}};

% Label
\node[draw=none, myred, anchor=north east]
    at (axis cs:-0.70,0.70)
    {\scalebox{0.65}{incident ray}};

\begin{scope}[shift={(axis direction cs:1,0)}]

    % Points
    \coordinate (I)  at (axis cs:-1.75,   0.5);
    \coordinate (P0) at (axis cs:-1.5, 0.0);
    \coordinate (P1) at (axis cs:-1.25, -0.5);
    \coordinate (P2) at (axis cs:-0.7, -0.3);
    \coordinate (P3) at (axis cs:-0.12,  -0.85);
    \coordinate (P4) at (axis cs:-0.0,   0.5);

    % Incident dashed ray
    \draw[myred, densely dashed] (I) -- (P0);

    % Segments + accolades + labels
    \raysegment{P0}{P1}
    \raysegment{P1}{P2}
    \raysegment{P2}{P3}
    \raysegment{P3}{P4}

    \node [draw = none, myred, anchor = north west] at (axis cs:-1.83,-0.28) {\scalebox{0.65}{$t_1 \w_1$}};
    \node [draw = none, myred, anchor = north west] at (axis cs:-1.1,-0.45) {\scalebox{0.65}{$t_2  \w_2$}};
    \node [draw = none, myred, anchor = north west] at (axis cs:-0.81,-0.59) {\scalebox{0.65}{$t_3  \w_3$}};
    \node [draw = none, myred, anchor = north west] at (axis cs:+0.02,-0.08) {\scalebox{0.65}{$t_4  \w_4$}};

    \node [draw = none, myblue, anchor = north, yshift=+1mm, xshift=-0.5mm] at (P1) {\scalebox{0.4}{\comicneue PAF!}};
    \node [draw = none, myblue, anchor = south, yshift=-1mm] at (P2) {\scalebox{0.4}{\comicneue PAF!}};
    \node [draw = none, myblue, anchor = north, yshift=+1mm] at (P3) {\scalebox{0.4}{\comicneue PAF!}};
    
\end{scope}

\iffalse
\begin{scope}[shift={(axis direction cs:1.125,0)}]
    % Points
    \coordinate (I)  at (axis cs:-1.75,   0.5);
    \coordinate (P0) at (axis cs:-1.5, 0.0);
    \coordinate (P1) at (axis cs:-1.37, -0.25);
    \coordinate (P2) at (axis cs:-0.7, 0.5);

    % Incident dashed ray
    \draw[myred, densely dashed] (I) -- (P0);
    \draw [->,>=stealth, myred] (P0) -- (P1) -- (P2);
    
\end{scope}

\begin{scope}[shift={(axis direction cs:1.0-0.125,0)}]
    % Points
    \coordinate (I)  at (axis cs:-1.75,   0.5);
    \coordinate (P0) at (axis cs:-1.5, 0.0);
    \coordinate (P1) at (axis cs:-1.43, -0.125);
    \coordinate (P2) at (axis cs:-1.7, -0.35);
    \coordinate (P3) at (axis cs:-1.3, 0.3);

    % Incident dashed ray
    \draw[myred, densely dashed] (I) -- (P0);
    \draw [->,>=stealth, myred] (P0) -- (P1) -- (P2) -- (P3);
    
\end{scope}
\fi 
\end{axis}
\end{tikzpicture}}
    \end{minipage}\hfill
    \begin{minipage}[t]{0.52\linewidth}
        \vspace{3pt}
        \setlength{\intextsep}{0pt}
        \begin{algorithm}[H]
            \caption{Random walk in a semi-infinite volume}
            \label{alg_walk}
            \scalebox{0.7}{%
            \begin{minipage}{\dimexpr\linewidth*10/7\relax}
            \begin{algorithmic}[1]
                \Function{StochasticRandomWalk}{$\w_1$, $\bn$}
                \State $\w$ $\gets$ $\w_1$
                \State $t$ $\gets$ \Call{DistanceToNextCollision}{$\w$}
                \State $(x, y, z)$ $\gets$ $t \, \w$
                \While{$[\bn \cdot (x, y, z)] < 0$}
                \State $\w$ $\gets$ \Call{DeviateRay}{$\w$}
                \State $t$ $\gets$ \Call{DistanceToNextCollision}{$\w$}
                \State $(x, y, z)$ $\gets$ $(x, y, z) + t \, \w$
                \EndWhile
                \State \Return $\w$
                \EndFunction
            \end{algorithmic}
            \end{minipage}%
            }
        \end{algorithm}
    \end{minipage}
    \caption{ \label{fig_geometry} Random-walk within a semi-infinite volume. }
\end{figure}
% =============================================================================================

% =============================================================================================
\paragraph{Stochasticity and Link to the BRDF}
As long as the collision and deviation laws ensure that the random walk exits the medium with probability one, 
repeated executions for a fixed input direction $\w_1$ define a normalized probability density $\mathcal R(\w_1,\w_o)$ over the upper hemisphere.
This exit density determines an energy-preserving BRDF through
\begin{equation}
    \mathcal R(\w_1, \w_o)
    =
    f_r(-\w_1,\w_o)\,z_o,
    \qquad
    z_o:=\w_o\cdot\bn.
    \label{eq:brdf-link}
\end{equation}
Here, $z_o$ is the usual cosine factor relating a BRDF to its directional density, while $-\w_1$ accounts for the conventional orientation of the incident BRDF direction.
Constructing the BRDF therefore amounts to specifying the probability laws governing the random walk.
In Algorithm~\ref{alg_walk}, these laws are implemented by \Call{DistanceToNextCollision}{} and \Call{DeviateRay}{}, 
which respectively determine the free-flight distances and the directional deviations produced by collisions.
Next, we use microflake theory to construct both functions and derive their probability distributions from the microflake normal distribution function.

% =============================================================================================
\paragraph{Microflake Model}
We now assume that the semi-infinite medium is a homogeneous specular microflake 
medium~\cite{jakob2010radiative}. 
Under this assumption, the ray travels freely until it collides with a microflake, after which it deviates based on 
the way the flake reflects incident rays. Both of these stochastic processes depend 
on the normal distribution function (NDF), which describes how the microflakes are statistically 
oriented throughout the medium via their normal $\w_m$. 
Letting $D := D(\w_m)$ denote the NDF, collisions occur according to the Beer-Lambert law, 
i.e., the distance to the next microflake collision follows the PDF
\begin{equation}
    \label{eq_free_flight}
    p_{\text{\scalebox{0.7}{\comicneue PAF!}}}(t \,|\, \w_j)
    =
    \sigma_j \, \exp \left(-\sigma_j \, t\right),
\end{equation}
where $\sigma_j$ denotes the projected area of the microflake distribution along direction $\w_j$
\begin{equation} 
    \sigma_j := \sigma(\w_j) = \int_{\mathcal S^2} (-\w_j \cdot \w_m)_+ \, D(\w_m) \, \dd \w_m.
    \label{eq:generic-extinction}
\end{equation} 
Once the collision point is determined, the ray's direction $\w_j \in \mathcal S^2$ 
deviates towards $\w_{j+1} \in \mathcal S^2$. This deviation is done according to the microflake specular 
phase function, which is also a PDF over deviated directions
\begin{equation}
    \label{eq_phase}
\Psi(\w_{j}, \w_{j+1}) = 
    \frac{D(\bh)}{4 \, \sigma_j}, \quad \bh:= \frac{\w_{j+1} - \w_{j}}{\|\w_{j+1} - \w_{j}\|}.
\end{equation}
We emphasize that Equations~\eqref{eq_free_flight} and~\eqref{eq_phase} are solely parameterized by the NDF.
The NDF thus fully determines the stochastic transport that occurs within the medium.
As long as $\sigma > 0$ for any downwelling direction, the random walks are guaranteed to escape the 
medium.

% =============================================================================================
% =============================================================================================
\subsection{Vertical-Depth Parameterization and First Two Scattering Orders}
\label{sec:generic:parameterization}

\begin{figure}[h]
    \pgfplotsset{every axis/.append style = {
    x = 1.75cm,
    y = 1.75cm,
    xmin = -2, xmax = 2,
    ymin = -1.5, ymax = 1,
    grid = both,
    axis line style = {color = black},
    minor grid style={dotted},
    major grid style={dotted},
    xticklabels = {},
    yticklabels = {},
    y label style = {rotate = -90},
    minor tick num = 3,
    ytick = {1,0,-1}
}}
\centering
\begin{tikzpicture}
\newcommand{\raysegment}[2]{%
    % segment 
    \draw[myred, ->, >=stealth] (#1) -- (#2);
    
    % bracket
    \draw[myred, decorate, opacity=0.4,
          decoration={brace, raise=3pt,
                      pre=moveto, post=moveto,
                      pre length=4pt, post length=3pt}]
    (#2) -- (#1);
}

\pgfmathsetseed{12346}
\begin{axis}[
    no markers,
    name = myaxis
]
% macroscopic medium
\addplot+[fill, color = myblue, opacity = 0.2] coordinates
{(-3,0) (3,0) (3,-3) (-3,-3)} --cycle;
\addplot [myblue] {0};
\node [opacity = 0.4, draw = none, myblue, anchor = north east] at (axis cs:-0.75,0.20) {\scalebox{0.65}{medium interface}};

% Label
\node[draw=none, myred, anchor=north east, opacity = 0.4]
    at (axis cs:-0.70,0.70)
    {\scalebox{0.65}{incident ray}};

\begin{scope}[shift={(axis direction cs:1,0)}]

    % Points
    \coordinate (I)  at (axis cs:-1.75,   0.5);
    \coordinate (P0) at (axis cs:-1.5, 0.0);
    \coordinate (P1) at (axis cs:-1.25, -0.5);
    \coordinate (P2) at (axis cs:-0.7, -0.3);
    \coordinate (P3) at (axis cs:-0.12,  -0.85);
    \coordinate (P4) at (axis cs:-0.0,   0.5);
    \coordinate (H1) at (axis cs:-1.5, -0.5);
    \coordinate (S0) at (axis cs:-2.5, 0.0);
    \coordinate (S1) at (axis cs:-2.5, -0.5);
    \coordinate (S2) at (axis cs:-2.25, -0.5);
    \coordinate (S3) at (axis cs:-2.25, -0.3);
    \coordinate (S4) at (axis cs:-2.0, -0.3);
    \coordinate (S5) at (axis cs:-2.0, -0.85);
    \coordinate (S6) at (axis cs:0.5, -0.85);
    \coordinate (S7) at (axis cs:0.5, 0.5);

    % Vertical-depth projection of the first segment
    %\path[fill=black, opacity=0.20]
    %    (S0) -- (P0) -- (P1) -- (H1) -- (S1) -- cycle;
    \draw[black, opacity=0.25, densely dashed]
        (S1) -- (P1);
    \draw[black, opacity=1.0, <->, >=stealth]
        (S0) -- node[left, xshift=0.7mm] {\scalebox{0.65}{$\Delta_1$}} (S1);

    % Vertical-depth projection of the second segment
    %\path[fill=black, opacity=0.20]
    %    (S2) -- (P1) -- (P2) -- (S3) -- cycle;
    \draw[black, opacity=0.25, densely dashed]
        (S3) -- (P2);
    \draw[black, opacity=1.0, <->, >=stealth]
        (S2) -- node[left, xshift=0.7mm] {\scalebox{0.65}{$\Delta_2$}} (S3);

    % Vertical-depth projection of the fourth segment
    %\path[fill=black, opacity=0.20]
    %    (P3) -- (S6) -- (S7) -- (P4) -- cycle;
    \draw[black, opacity=0.25, densely dashed]
        (P4) -- (S7);
    %\draw[black, opacity=1.0, densely dashed]
    %    (P3) -- (S6);
    \draw[black, opacity=1.0, <->, >=stealth]
        (S6) -- node[left, xshift=4.5mm] {\scalebox{0.65}{$\Delta_4$}} (S7);

    % Vertical-depth projection of the third segment
    \draw[black, opacity=0.25, densely dashed]
        (S5) -- (S6);
    \draw[black, opacity=1.0, <->, >=stealth]
        (S4) -- node[left, xshift=0.9mm] {\scalebox{0.65}{$\Delta_3$}} (S5);

    % Incident dashed ray
    \draw[myred, densely dashed, opacity = 0.4] (I) -- (P0);

    % Segments + accolades + labels
    \raysegment{P0}{P1}
    \raysegment{P1}{P2}
    \raysegment{P2}{P3}
    \raysegment{P3}{P4}

    \node [draw = none, myred, opacity = 0.4, anchor = north west] at (axis cs:-1.83,-0.28) {\scalebox{0.65}{$t_1 \w_1$}};
    \node [draw = none, myred, opacity = 0.4, anchor = north west] at (axis cs:-1.1,-0.45) {\scalebox{0.65}{$t_2  \w_2$}};
    \node [draw = none, myred, opacity = 0.4, anchor = north west] at (axis cs:-0.81,-0.59) {\scalebox{0.65}{$t_3  \w_3$}};
    \node [draw = none, myred, opacity = 0.4, anchor = north west] at (axis cs:+0.02,-0.08) {\scalebox{0.65}{$t_4  \w_4$}};

    \node [opacity = 0.4, draw = none, myblue, anchor = north, yshift=+1mm, xshift=-0.5mm] at (P1) {\scalebox{0.4}{\comicneue PAF!}};
    \node [opacity = 0.4, draw = none, myblue, anchor = south, yshift=-1mm] at (P2) {\scalebox{0.4}{\comicneue PAF!}};
    \node [opacity = 0.4, draw = none, myblue, anchor = north, yshift=+1mm] at (P3) {\scalebox{0.4}{\comicneue PAF!}};
    
\end{scope}

\iffalse
\begin{scope}[shift={(axis direction cs:1.125,0)}]
    % Points
    \coordinate (I)  at (axis cs:-1.75,   0.5);
    \coordinate (P0) at (axis cs:-1.5, 0.0);
    \coordinate (P1) at (axis cs:-1.37, -0.25);
    \coordinate (P2) at (axis cs:-0.7, 0.5);

    % Incident dashed ray
    \draw[myred, densely dashed] (I) -- (P0);
    \draw [->,>=stealth, myred] (P0) -- (P1) -- (P2);
    
\end{scope}

\begin{scope}[shift={(axis direction cs:1.0-0.125,0)}]
    % Points
    \coordinate (I)  at (axis cs:-1.75,   0.5);
    \coordinate (P0) at (axis cs:-1.5, 0.0);
    \coordinate (P1) at (axis cs:-1.43, -0.125);
    \coordinate (P2) at (axis cs:-1.7, -0.35);
    \coordinate (P3) at (axis cs:-1.3, 0.3);

    % Incident dashed ray
    \draw[myred, densely dashed] (I) -- (P0);
    \draw [->,>=stealth, myred] (P0) -- (P1) -- (P2) -- (P3);
    
\end{scope}
\fi 
\end{axis}
\end{tikzpicture}
    \caption{ \label{fig:depth-parameterization} Depth parameterization for random walks within the medium. }
\end{figure}
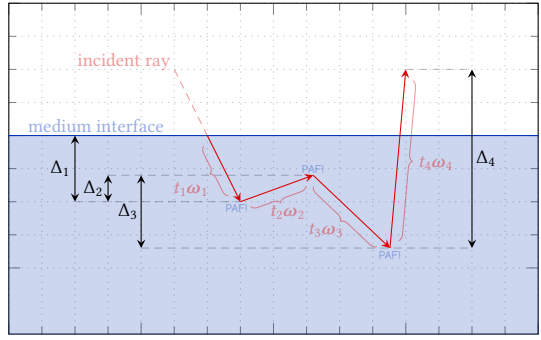

% =============================================================================================
\paragraph{Depth Parameterization}
Equations~\eqref{eq_free_flight} and~\eqref{eq_phase} describe the random walk in three dimensions. Since the medium is 
homogeneous parallel to its boundary, lateral collision coordinates have no influence on whether a ray escapes. 
Therefore, similarly to Bitterli and d'Eon~\cite{bitterli2022positionfree}, we only track the vertical depth of the walk.
For a ray traveling along $\w_j$, the projected area of Equation~\eqref{eq:generic-extinction} gives the collision rate per unit 
vertical depth
  \begin{equation}
      \kappa_j
      :=
      \kappa(\w_j)
      =
      \frac{\sigma_j}{|z_j|}.
      \label{eq:generic-kappa}
  \end{equation}
Consequently, the vertical distance $\Delta\geq0$ to the next potential collision is exponentially distributed:
  \begin{equation}
      p_j(\Delta)
      :=
      p(\Delta\mid\w_j)
      =
      \kappa_j e^{-\kappa_j\Delta}.
      \label{eq:generic-free-flight}
  \end{equation}
This one-dimensional reduction allows us to integrate the collision depths analytically for the first two scattering orders.

\begin{figure}[h]
    \centering
    \begin{tabular}{@{}ccc@{}}
        {\pgfplotsset{every axis/.append style = {
    x = 1.75cm,
    y = 1.75cm,
    xmin = -1.25, xmax = 0.75,
    ymin = -1.5, ymax = 1,
    grid = both,
    axis line style = {color = black},
    minor grid style={dotted},
    major grid style={dotted},
    xticklabels = {},
    yticklabels = {},
    y label style = {rotate = -90},
    minor tick num = 1,
    xtick = {1,0,-1},
    ytick = {1,0,-1}
}}
\centering
\begin{tikzpicture}
\newcommand{\raysegment}[2]{%
    % segment 
    \draw[myred, ->, >=stealth] (#1) -- (#2);
    
    % bracket
    \draw[myred, decorate, opacity=0.4,
          decoration={brace, raise=3pt,
                      pre=moveto, post=moveto,
                      pre length=4pt, post length=3pt}]
    (#2) -- (#1);
}

\pgfmathsetseed{12346}
\begin{axis}[
    no markers,
    name = myaxis
]
% macroscopic medium
\addplot+[fill, color = myblue, opacity = 0.2] coordinates
{(-3,0) (3,0) (3,-3) (-3,-3)} --cycle;
\addplot [myblue] {0};

\begin{scope}[shift={(axis direction cs:1,0)}]

    % Points
    \coordinate (I)  at (axis cs:-1.75,   0.5);
    \coordinate (P0) at (axis cs:-1.5, 0.0);
    \coordinate (P1) at (axis cs:-1.25, -0.5);
    \coordinate (P2) at (axis cs:-0.7, -0.3);
    \coordinate (P3) at (axis cs:-0.12,  -0.85);
    \coordinate (P4) at (axis cs:-1.05,   0.5);
    \coordinate (H1) at (axis cs:-1.5, -0.5);
    \coordinate (S0) at (axis cs:-1.85, 0.0);
    \coordinate (S1) at (axis cs:-1.85, -0.5);
    \coordinate (S2) at (axis cs:-0.7, -0.5);
    \coordinate (S3) at (axis cs:-0.7, +0.5);

    % Vertical-depth projection of the first segment
    %\path[fill=black, opacity=0.20]
    %    (S0) -- (P0) -- (P1) -- (H1) -- (S1) -- cycle;
    \draw[black, opacity=0.25, densely dashed]
        (S1) -- (P1);
    \draw[black, opacity=1.0, ->, >=stealth]
        (S0) -- node[left, xshift=0.7mm] {\scalebox{0.65}{$-\Delta_1$}} (S1);

    \draw[black, opacity=0.25, densely dashed]
        (S2) -- (P1);
    \draw[black, opacity=0.25, densely dashed]
        (S3) -- (P4);
    \draw[black, opacity=1.0, ->, >=stealth]
        (S2) -- node[left, xshift=6.2mm, yshift=1.5mm] {\scalebox{0.65}{$+\Delta_2$}} (S3);

    % Incident dashed ray
    \draw[myred, densely dashed, opacity = 0.4] (I) -- (P0);

    % Segments + accolades + labels
    \raysegment{P0}{P1}
    \raysegment{P1}{P4}

    \node [draw = none, myred, opacity = 0.4, anchor = north west] at (axis cs:-1.83,-0.22) {\scalebox{0.65}{$t_1 \w_1$}};
    \node [draw = none, myred, opacity = 0.4, anchor = north west] at (axis cs:-1.1,+0.05) {\scalebox{0.65}{$t_2  \w_2$}};
    \node [opacity = 0.4, draw = none, myblue, anchor = north, yshift=+0.5mm, xshift=-0.1mm] at (P1) {\scalebox{0.4}{\comicneue PAF!}};
    \node [draw = none, black, opacity = 1.0, anchor = north] at (axis cs:-1.25,0.95) {\scalebox{0.8}{$\boxed{\Delta_1 \leq \Delta_2}$}};
   
\end{scope}

\end{axis}
\end{tikzpicture}}
        &
        {\centering
\begin{tikzpicture}
\newcommand{\raysegment}[2]{%
    % segment
    \draw[myred, ->, >=stealth] (#1) -- (#2);

    % bracket
    \draw[myred, decorate, opacity=0.4,
          decoration={brace, raise=3pt,
                      pre=moveto, post=moveto,
                      pre length=4pt, post length=3pt}]
    (#2) -- (#1);
}

\begin{axis}[
    x = 1.75cm,
    y = 1.75cm,
    xmin = -2.25, xmax = 0.25,
    ymin = -1.5, ymax = 1,
    grid = both,
    axis line style = {color = black},
    minor grid style={dotted},
    major grid style={dotted},
    xticklabels = {},
    yticklabels = {},
    y label style = {rotate = -90},
    minor tick num = 1,
    xtick = {1,0,-1,-2},
    ytick = {1,0,-1,-2}
]
    % Semi-infinite medium
    \addplot[fill, color=myblue, opacity=0.2] coordinates
        {(-4,0) (4,0) (4,-4) (-4,-4)} -- cycle;
    \addplot[myblue, domain=-4:4] {0};

    % Path vertices: down, down, up
    \coordinate (I)  at (axis cs:-1.75,   0.5);
    \coordinate (P0) at (axis cs:-1.5, 0.0);
    \coordinate (P1) at (axis cs:-1.25, -0.5);
    \coordinate (P2) at (axis cs:-0.85,-0.9);
    \coordinate (P3) at (axis cs:-0.58,0.55);

    % Vertical-depth parameterization
    \coordinate (D1T) at (axis cs:-1.91,0.0);
    \coordinate (D1B) at (axis cs:-1.91,-0.5);
    \coordinate (D2T) at (axis cs:-1.91,-0.5);
    \coordinate (D2B) at (axis cs:-1.91,-0.9);
    \coordinate (D3B) at (axis cs:-0.2,-0.9);
    \coordinate (D3T) at (axis cs:-0.2,0.55);

    \draw[black, densely dashed, opacity=0.25] (D1B) -- (P1);
    \draw[black, ->, >=stealth]
        (D1T) -- node[left, xshift=0.7mm] {\scalebox{0.65}{$-\Delta_1$}} (D1B);
    \draw[black, densely dashed, opacity=0.25] (D2B) -- (P2);
    \draw[black, ->, >=stealth]
        (D2T) -- node[left, xshift=0.7mm] {\scalebox{0.65}{$-\Delta_2$}} (D2B);
    \draw[black, densely dashed, opacity=0.25] (P2) -- (D3B);
    \draw[black, densely dashed, opacity=0.25] (P3) -- (D3T);
    \draw[black, ->, >=stealth]
        (D3B) -- node[left, xshift=6mm] {\scalebox{0.65}{$+\Delta_3$}} (D3T);

    % Double-scattering path
    \draw[myred, densely dashed, opacity = 0.4] (I) -- (P0);
    \raysegment{P0}{P1}
    \raysegment{P1}{P2}
    \raysegment{P2}{P3}

    \node[draw=none, myred, opacity=0.4, anchor=north west]
        at (axis cs:-1.8,-0.22) {\scalebox{0.65}{$t_1\w_1$}};
    \node[draw=none, myred, opacity=0.4, anchor=north west]
        at (axis cs:-1.40,-0.7) {\scalebox{0.65}{$t_2\w_2$}};
    \node[draw=none, myred, opacity=0.4, anchor=west]
        at (axis cs:-0.66,-0.2) {\scalebox{0.65}{$t_3\w_3$}};
    \node[draw=none, black, anchor=north]
        at (axis cs:-1.05, 0.95)
        {\scalebox{0.8}{$\boxed{\Delta_1+\Delta_2\leq\Delta_3}$}};

    \node [opacity = 0.4, draw = none, myblue, anchor = north, yshift=+0.5mm, xshift=-0.1mm] at (P1) {\scalebox{0.4}{\comicneue PAF!}};
    \node [opacity = 0.4, draw = none, myblue, anchor = north, yshift=+0.5mm, xshift=-0.1mm] at (P2) {\scalebox{0.4}{\comicneue PAF!}};
\end{axis}
\end{tikzpicture}}
        &
        {\centering
\begin{tikzpicture}
\newcommand{\raysegment}[2]{%
    % segment
    \draw[myred, ->, >=stealth] (#1) -- (#2);

    % bracket
    \draw[myred, decorate, opacity=0.4,
          decoration={brace, raise=3pt,
                      pre=moveto, post=moveto,
                      pre length=4pt, post length=3pt}]
    (#2) -- (#1);
}

\begin{axis}[
    x = 1.75cm,
    y = 1.75cm,
    xmin = -2.25, xmax = 0.25,
    ymin = -1.5, ymax = 1,
    grid = both,
    axis line style = {color = black},
    minor grid style={dotted},
    major grid style={dotted},
    xticklabels = {},
    yticklabels = {},
    y label style = {rotate = -90},
    minor tick num = 1,
    xtick = {1,0,-1,-2},
    ytick = {1,0,-1,-2}
]
    % Semi-infinite medium
    \addplot[fill, color=myblue, opacity=0.2] coordinates
        {(-4,0) (4,0) (4,-4) (-4,-4)} -- cycle;
    \addplot[myblue, domain=-4:4] {0};

    % Path vertices: down, up, up
    \coordinate (I)  at (axis cs:-1.75,0.5);
    \coordinate (P0) at (axis cs:-1.5,0.0);
    \coordinate (P1) at (axis cs:-1.15,-0.7);
    \coordinate (P2) at (axis cs:-0.6,-0.5);
    \coordinate (P3) at (axis cs:-0.5,0.4);

    % Vertical-depth parameterization
    \coordinate (D1T) at (axis cs:-1.91,0.0);
    \coordinate (D1B) at (axis cs:-1.91,-0.7);
    \coordinate (D2B) at (axis cs:-0.1,-0.7);
    \coordinate (D2T) at (axis cs:-0.1,-0.5);
    \coordinate (D3B) at (axis cs:-0.1,-0.5);
    \coordinate (D3T) at (axis cs:-0.1,0.4);

    \draw[black, densely dashed, opacity=0.25] (D1B) -- (P1);
    \draw[black, ->, >=stealth]
        (D1T) -- node[left, xshift=0.7mm] {\scalebox{0.65}{$-\Delta_1$}} (D1B);
    \draw[black, densely dashed, opacity=0.25] (D2B) -- (P1);
    \draw[black, densely dashed, opacity=0.25] (D2T) -- (P2);
    \draw[black, ->, >=stealth]
        (D2B) -- node[left, xshift=6mm] {\scalebox{0.65}{$+\Delta_2$}} (D2T);
    \draw[black, densely dashed, opacity=0.25] (P3) -- (D3T);
    \draw[black, ->, >=stealth]
        (D3B) -- node[left, xshift=6mm, yshift=-0.5mm] {\scalebox{0.65}{$+\Delta_3$}} (D3T);

    % Double-scattering path
    \draw[myred, densely dashed, opacity = 0.4] (I) -- (P0);
    \raysegment{P0}{P1}
    \raysegment{P1}{P2}
    \raysegment{P2}{P3}

    \node[draw=none, myred, opacity=0.4, anchor=north west]
        at (axis cs:-1.75,-0.30) {\scalebox{0.65}{$t_1\w_1$}};
    \node[draw=none, myred, opacity=0.4, anchor=north west]
        at (axis cs:-0.95,-0.65) {\scalebox{0.65}{$t_2\w_2$}};
    \node[draw=none, myred, opacity=0.4, anchor=west]
        at (axis cs:-0.5,-0.1) {\scalebox{0.65}{$t_3\w_3$}};
    \node[draw=none, black, anchor=north]
        at (axis cs:-1.05,0.95)
        {\scalebox{0.8}{$\boxed{\Delta_2\leq\Delta_1\leq\Delta_2+\Delta_3}$}};

    \node[opacity=0.4, draw=none, myblue, anchor=north, yshift=+0.5mm, xshift=-0.1mm]
        at (P1) {\scalebox{0.4}{\comicneue PAF!}};
    \node[opacity=0.4, draw=none, myblue, anchor=north, yshift=+0.5mm, xshift=+0.8mm]
        at (P2) {\scalebox{0.4}{\comicneue PAF!}};
\end{axis}
\end{tikzpicture}}
        \\[-0.25em]
        (a) & (b) & (c)
    \end{tabular}
    \caption{ \label{fig:scattering} All possible path configurations for escaping the medium after (a) one collision and 
    (b, c) two collisions. }
\end{figure}
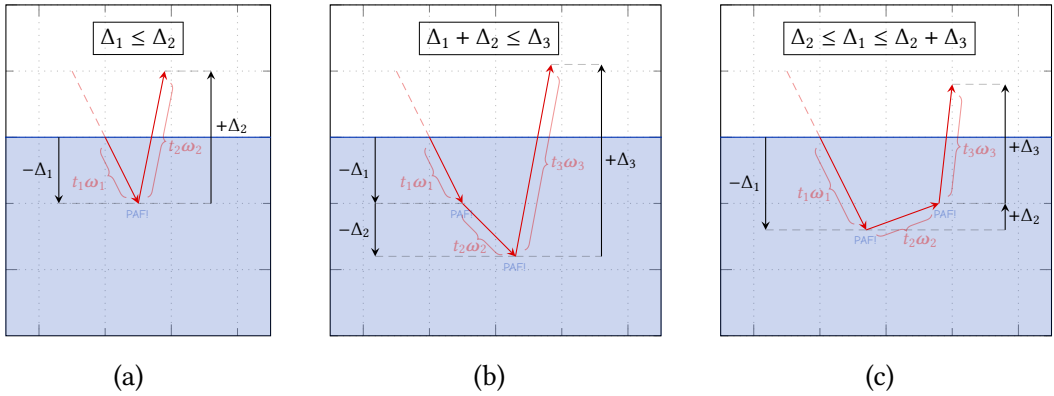

% =============================================================================================
\newpage
\paragraph{Escape Probability After One Collision}
Consider an incident direction $\w_1$ and an escaping direction $\w_2$.
If the first collision occurs at depth $\Delta_1$, the outgoing segment must then travel
at least the same vertical distance without another collision. 
The probability of escaping the medium after exactly one collision is therefore
\begin{align}
P_1(\w_1,\w_2)
&=
    \underbrace{
        \vphantom{\int_{\Delta_1}^\infty p_2(\Delta_2) \, \dd \Delta_2}
        \int_0^\infty p_1(\Delta_1)
    }_{
        \scalebox{0.6}{\text{move down by $\Delta_1$}}
    }
    \,
    \underbrace{
        \int_{\Delta_1}^\infty p_2(\Delta_2) \, 
    }_{
        \scalebox{0.6}{\text{now up by at least $\Delta_1$}}
    }
    \dd \Delta_2 \, \dd \Delta_1 \notag\\
&=
        \int_0^\infty \kappa_1 \, e^{-\kappa_1 \, \Delta_1}
        \left[\int_{\Delta_1}^\infty \kappa_2 \, e^{-\kappa_2 \, \Delta_2} \, \dd \Delta_2\right]
    \dd \Delta_1 \notag\\
&=\int_0^\infty \kappa_1 \, e^{-(\kappa_1+\kappa_2) \, \Delta_1} \,\dd \Delta_1 \notag \\
&=\frac{\kappa_1}{\kappa_1+\kappa_2}.
\label{eq:P1}
\end{align}
% =============================================================================================
\paragraph{Single-Scattering Exit-Direction Density}
Multiplying the single-collision escape probability by the phase function gives the
single-scattering contribution to the exit-direction density
\begin{equation}
    \mathcal{R}_1(\w_1,\w_2)
    =\frac{\kappa_1}{\kappa_1+\kappa_2} \, \Psi(\w_1, \w_2).
    \label{eq:generic-single}
\end{equation}
Since both $\kappa$ and $\Psi$ are determined by the NDF, so is this single-scattering exit-direction density.

% =============================================================================================
\paragraph{Escape Probability After Two Collisions}
An escaping path with two scattering events has two possible configurations, depending on the sign of the intermediate direction $\w_2$, as illustrated in Figure~\ref{fig:scattering}.
If the intermediate direction $\w_2$ is downwelling, then the ray must escape from the accumulated depth:
\begin{align}
P_2(\w_1,\w_2,\w_3)\big|_{z_2<0}
&=
\underbrace{
    \vphantom{\int_{\Delta_1+\Delta_2}^\infty p_3(\Delta_3)}
    \int_0^\infty p_1(\Delta_1)
}_{
    \scalebox{0.6}{\text{move down by $\Delta_1$}}
}
\underbrace{
    \vphantom{\int_{\Delta_1+\Delta_2}^\infty p_3(\Delta_3)}
    \int_0^\infty p_2(\Delta_2)
}_{
    \scalebox{0.6}{\text{further down by $\Delta_2$}}
}
\underbrace{
    \int_{\Delta_1+\Delta_2}^\infty p_3(\Delta_3)
}_{
    \scalebox{0.6}{\text{now up by $\Delta_3 \geq \Delta_1 + \Delta_2$}}
}
  \dd \Delta_3 \, \dd \Delta_2\, \dd \Delta_1 \notag\\
&=
\frac{\kappa_1}{\kappa_1+\kappa_3}
\frac{\kappa_2}{\kappa_2+\kappa_3}.
\label{eq:P2-down}
\end{align}
If instead the intermediate direction $\w_2$ is upwelling, it must collide before reaching the boundary so that the final ray can escape:
\begin{align}
P_2(\w_1,\w_2,\w_3)\big|_{z_2>0}
&=
\underbrace{
    \vphantom{\int_{\Delta_1-\Delta_2}^\infty p_3(\Delta_3)}
    \int_0^\infty p_1(\Delta_1)
}_{
    \scalebox{0.6}{\text{move down by $\Delta_1$}}
}
\underbrace{
    \vphantom{\int_{\Delta_1-\Delta_2}^\infty p_3(\Delta_3)}
    \int_0^{\Delta_1} p_2(\Delta_2)
}_{
    \scalebox{0.6}{\text{now up by $\Delta_2 \leq \Delta_1$}}
}
\underbrace{
    \int_{\Delta_1-\Delta_2}^\infty p_3(\Delta_3)
}_{
    \scalebox{0.6}{\text{further up by $\Delta_3 \geq \Delta_1 - \Delta_2$}}
}
  \dd \Delta_3 \, \dd \Delta_2\, \dd \Delta_1 \notag\\
&=
\frac{\kappa_1}{\kappa_1+\kappa_3}
\frac{\kappa_2}{\kappa_1+\kappa_2}.
\label{eq:P2-up}
\end{align}
Using the single-scattering escape probability to combine both cases gives
\begin{equation}
P_2(\w_1,\w_2,\w_3)
=
\frac{
\ind_{z_2<0}\,\kappa_1\, P_1(\w_2,\w_3)
+
\ind_{z_2>0}\,\kappa_2 \, P_1(\w_1,\w_2)
}{\kappa_1+\kappa_3}.
\label{eq:P2}
\end{equation}

% =============================================================================================
\paragraph{Double-Scattering Exit-Direction Density}
Marginalizing the resulting path density over the intermediate direction gives the double-scattering contribution to the exit-direction density
\begin{equation}
\mathcal{R}_2(\w_1,\w_3)
=
\int_{\mathcal S^2}
P_2(\w_1,\w_2,\w_3)
\,
\Psi(\w_1,\w_2)
\,
\Psi(\w_2,\w_3)
\,
\dd\w_2.
\label{eq:generic-double}
\end{equation}

% =============================================================================================
% =============================================================================================
\subsection{Optional -- Isotropic Scattering Orders}
\label{sec:generic:optional}

The derivation in this subsection is not required for the rest of the paper. Its goal is to illustrate 
how the framework developed so far efficiently recovers known results.

\paragraph*{Isotropic Scattering}
We apply it to the simplest case of isotropic scattering, obtained from a constant microflake NDF:
%The derivation in this subsection is not required for the rest of the paper, but illustrates how the framework developed so far 
%recovers known results efficiently. We instantiate our derivations for the simplest case of isotropic scattering. 
%This is accomplished using a constant microflake NDF:
  \begin{equation}
      D(\w_m) = \frac{1}{4\pi} \quad \Rightarrow \quad  \Psi(\w_j,\w_{j+1})=\frac{1}{4\pi}, \quad \kappa(\w_j)=\frac{1}{4|z_j|}.
      \label{eq:chandrasekhar-ndf}
  \end{equation}
We now derive the first two scattering orders of the corresponding BRDF.

% =============================================================================================
\paragraph*{Single-scattering BRDF}
Inserting Equation~\eqref{eq:chandrasekhar-ndf} into Equation~\eqref{eq:generic-single} gives
  \begin{equation}
      \mathcal R_1(\w_1,\w_o)
      =
      \frac{1}{4\pi}
      \frac{z_o}{z_o-z_1}.
      \label{eq:chandrasekhar-R1}
  \end{equation}
We then retrieve the single-scattering BRDF from Equation~\eqref{eq:brdf-link}:
  \begin{equation}
      f_{r_1}(\w_i,\w_o)
      =
      \frac{\mathcal R_1(-\w_i, \w_o)}{z_o}
      =
      \frac{1}{4\pi}
      \frac{1}{z_o+z_i}.
      \label{eq:chandrasekhar-f1}
  \end{equation}

% =============================================================================================
\paragraph*{Double-scattering BRDF}
Inserting Equations~\eqref{eq:P2} and~\eqref{eq:chandrasekhar-ndf} into Equation~\eqref{eq:generic-double} gives
  \begin{align}
  \mathcal R_2(\w_1, \w_o)
  &=
  \frac{1}{(4\pi)^2}
  \int_{\mathcal S^2}
  \frac{z_o}{z_o-z_1}
  \left[
  \ind_{z_2<0}
  \frac{z_o}{z_o-z_2}
  +
  \ind_{z_2>0}
  \frac{-z_1}{z_2-z_1}
  \right]
  \dd\w_2
  \notag\\
  &=
  \frac{1}{(4\pi)^2}
  \frac{z_o}{z_o-z_1}
  \int_0^{2\pi}
  \left[
  \int_{-1}^{0}
  \frac{z_o}{z_o-z_2}\,\dd z_2
  +
  \int_{0}^{1}
  \frac{-z_1}{z_2-z_1}\,\dd z_2
  \right]
  \dd\phi_2
  \notag\\
  &=
  \mathcal R_1(\w_1, \w_o)
  \, \frac{1}{2}
  \left[
  \int_{-1}^{0}
  \frac{z_o}{z_o-z_2}\,\dd z_2
  +
  \int_{0}^{1}
  \frac{-z_1}{z_2-z_1}\,\dd z_2
  \right]
  \notag\\
  &=
  \mathcal R_1(\w_1, \w_o)
  \, \frac{1}{2}
  \left[
  z_o\log\left(\frac{z_o+1}{z_o}\right)
  -
  z_1\log\left(\frac{z_1-1}{z_1}\right)
  \right].
  \label{eq:chandrasekhar-R2}
  \end{align}
We then retrieve the double-scattering BRDF from Equation~\eqref{eq:brdf-link}:
  \begin{align}
      f_{r_2}(\w_i,\w_o)
      &=
      \frac{\mathcal R_2(-\w_i, \w_o)}{z_o}
      \notag\\
      &=
      f_{r_1}(\w_i,\w_o) \,
      \frac{1}{2}
      \left[
      z_i\log\left(\frac{z_i+1}{z_i}\right)
      +
      z_o\log\left(\frac{z_o+1}{z_o}\right)
      \right]
      \notag\\
      &=
      \frac{1}{8\pi}
      \frac{
      z_i\log\left(\frac{z_i+1}{z_i}\right)
      +
      z_o\log\left(\frac{z_o+1}{z_o}\right)
      }{
      z_i+z_o
      }.
      \label{eq:chandrasekhar-f2}
  \end{align}
  Equations~\eqref{eq:chandrasekhar-f1} and~\eqref{eq:chandrasekhar-f2} are known: they correspond to the first 
  two scattering orders of the 
  Chandrasekhar BRDF~\cite[Sec.~43.4, p.~239]{deon2022hitchhikers}.
  The novelty here is that we were able to derive them straightforwardly from the geometry of the transport.  

% =============================================================================================
% =============================================================================================
% =============================================================================================

\section{Specialization to Heightfield-Like Transport}
\label{sec:heightfield}

We now specialize the semi-infinite microflake medium so that it produces heightfield-like transport.
We show that this particular configuration allows us to derive the escape probability for any scattering order
and express reflectance as a series based on this probability.

% =============================================================================================
% =============================================================================================
\subsection{One-Sided Microflake Distribution}

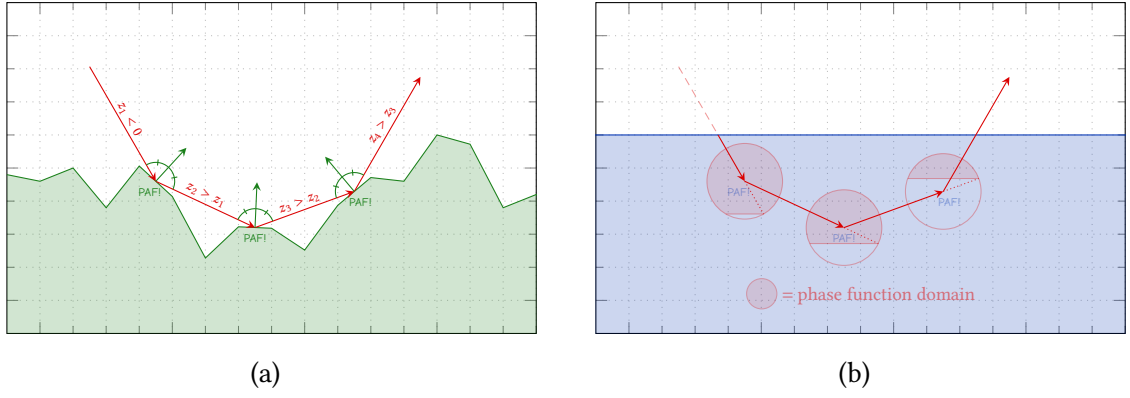
\begin{figure}[t]
    \centering
    \begin{tabular}{@{}cc@{}}
        {\begin{tikzpicture}
\begin{axis}[
    x = 1.75cm,
    y = 1.75cm,
    xmin = -2.25, xmax = 1.75,
    ymin = -1.25, ymax = 1.25,
    grid = both,
    axis line style = {color = black},
    minor grid style={dotted},
    major grid style={dotted},
    xticklabels = {},
    yticklabels = {},
    %tick style = {draw = none},
    minor tick num = 3,
    xtick = {-2,-1,0,1,2},
    ytick = {-1,0,1}
]
    % Coarsely discretized heightfield
    \addplot[draw=none, fill=mygreen, opacity=0.2] coordinates {
        (-2.25,-0.050)
        (-2.00,-0.100)
        (-1.75,-0.00)
        (-1.50, -0.300)
        (-1.25, 0.015)
        (-1.00,-0.215)
        (-0.75,-0.680)
        (-0.50,-0.444)
        (-0.25,-0.455)
        ( 0.00,-0.620)
        ( 0.25,-0.282)
        ( 0.50,-0.072)
        ( 0.75, -0.100)
        ( 1.00, 0.250)
        ( 1.25, 0.180)
        ( 1.50, -0.30)
        ( 1.75, -0.20)
        ( 2.00, -0.4)
        ( 2.00,-3.000)
        (-2.25,-3.000)
    } -- cycle;
    \addplot[mygreen] coordinates {
        (-2.25,-0.050)
        (-2.00,-0.100)
        (-1.75,-0.00)
        (-1.50, -0.300)
        (-1.25, 0.015)
        (-1.00,-0.215)
        (-0.75,-0.680)
        (-0.50,-0.444)
        (-0.25,-0.455)
        ( 0.00,-0.620)
        ( 0.25,-0.282)
        ( 0.50,-0.072)
        ( 0.75, -0.100)
        ( 1.00, 0.250)
        ( 1.25, 0.180)
        ( 1.50, -0.30)
        ( 1.75, -0.20)
        ( 2.00, -0.4)
        ( 2.00,-3.000)
        (-2.25,-3.000)
    };

    % Three-bounce specular path
    \coordinate (I)  at (axis cs:-1.625,0.766);
    \coordinate (P1) at (axis cs:-1.125,-0.100);
    \coordinate (P2) at (axis cs:-0.375,-0.450);
    \coordinate (P3) at (axis cs:0.375,-0.177);
    \coordinate (O)  at (axis cs:0.875,0.689);

    \draw[myred, ->, >=stealth] (I) -- (P1);
    \draw[myred, ->, >=stealth] (P1) -- (P2);
    \draw[myred, ->, >=stealth] (P2) -- (P3);
    \draw[myred, ->, >=stealth] (P3) -- (O);

    % Microfacet normals and equal-angle indicators
    \draw[mygreen, ->, >=stealth] (P1) -- ++(47.5:6mm);
        %node[draw=none, anchor=west] {\scalebox{0.55}{$\w_{m,1}$}};
    \draw[mygreen, ->, >=stealth] (P2) -- ++(87.5:6mm);
        %node[draw=none, anchor=south] {\scalebox{0.55}{$\w_{m,2}$}};
    \draw[mygreen, ->, >=stealth] (P3) -- ++(130:6mm);
        %node[draw=none, anchor=south east] {\scalebox{0.55}{$\w_{m,3}$}};

    \draw[mygreen] (P1) ++(-25:2.5mm) arc[start angle=-25, end angle=47.5, radius=2.5mm];
    \draw[mygreen] (P1) ++(47.5:2.5mm) arc[start angle=47.5, end angle=120, radius=2.5mm];
    \draw[mygreen] (P1) ++(11.25:2.1mm) -- ++(11.25:0.8mm);
    \draw[mygreen] (P1) ++(83.75:2.1mm) -- ++(83.75:0.8mm);

    \draw[mygreen] (P2) ++(20:2.5mm) arc[start angle=20, end angle=87.5, radius=2.5mm];
    \draw[mygreen] (P2) ++(87.5:2.5mm) arc[start angle=87.5, end angle=155, radius=2.5mm];
    \draw[mygreen] (P2) ++(53.75:2.1mm) -- ++(53.75:0.8mm);
    \draw[mygreen] (P2) ++(121.25:2.1mm) -- ++(121.25:0.8mm);

    \draw[mygreen] (P3) ++(60:2.5mm) arc[start angle=60, end angle=130, radius=2.5mm];
    \draw[mygreen] (P3) ++(130:2.5mm) arc[start angle=130, end angle=200, radius=2.5mm];
    \draw[mygreen] (P3) ++(95:2.1mm) -- ++(95:0.8mm);
    \draw[mygreen] (P3) ++(165:2.1mm) -- ++(165:0.8mm);

    \node[draw=none, myred, anchor=south east, rotate=-60]
        at (axis cs:-1.32,0.125) {\scalebox{0.5}{$z_1 < 0$}};
    \node[draw=none, myred, anchor=north, rotate=-25]
        at (axis cs:-0.70,-0.1) {\scalebox{0.5}{$z_2 > z_1$}};
    \node[draw=none, myred, anchor=south, rotate=20]
        at (axis cs:0.00,-0.37) {\scalebox{0.5}{$z_3 > z_2$}};
    \node[draw=none, myred, anchor=west, rotate=60]
        at (axis cs:0.47,0.125) {\scalebox{0.5}{$z_4 > z_3$}};

    \node[draw=none, mygreen, anchor=north, xshift=-0.8mm, yshift=0.5mm]
        at (P1) {\scalebox{0.4}{\comicneue PAF!}};
    \node[draw=none, mygreen, anchor=north, yshift=0.5mm]
        at (P2) {\scalebox{0.4}{\comicneue PAF!}};
    \node[draw=none, mygreen, anchor=north, xshift=+0.8mm, yshift=0.5mm]
        at (P3) {\scalebox{0.4}{\comicneue PAF!}};
\end{axis}
\end{tikzpicture}}
        &
        {\begin{tikzpicture}
\newcommand{\phasecap}[4]{%
    \path[draw=none, fill=myred, opacity=0.1]
        (#1) ++(#2:5mm)
        arc[start angle=#2, end angle=#3, radius=5mm] -- cycle;
    \draw[myred, opacity=0.3] (#1) circle[radius=5mm];
    \draw[myred, opacity=0.8, densely dotted] (#1) -- ++(#2:5mm);
    \draw[myred, opacity=0.3] (#1) ++(#3:5mm) -- ++(0:#4);
}

\begin{axis}[
    x = 1.75cm,
    y = 1.75cm,
    xmin = -2.25, xmax = 1.75,
    ymin = -1.25, ymax = 1.25,
    grid = both,
    axis line style = {color = black},
    minor grid style={dotted},
    major grid style={dotted},
    xticklabels = {},
    yticklabels = {},
    minor tick num = 3,
    xtick = {-2,-1,0,1,2},
    ytick = {-1,0,1}
]
    % Semi-infinite medium
    \addplot[draw=none, fill=myblue, opacity=0.2] coordinates {
        (-3,0.25)
        ( 2,0.25)
        ( 2,-2)
        (-3,-2)
    } -- cycle;
    \addplot[myblue, domain=-3:3] {0.25};

    % Same path as in heightfield.tex
    \coordinate (I)  at (axis cs:-1.625, 0.766);
    \coordinate (P0) at (axis cs:-1.327, 0.250);
    \coordinate (P1) at (axis cs:-1.125,-0.100);
    \coordinate (P2) at (axis cs:-0.375,-0.450);
    \coordinate (P3) at (axis cs: 0.375,-0.177);
    \coordinate (O)  at (axis cs: 0.875, 0.689);

    % Direction-space circles and spherical-cap chords
    \phasecap{P1}{-60}{240}{5mm}
    \phasecap{P2}{-25}{205}{9.06mm}
    \phasecap{P3}{20}{160}{9.40mm}

    % Random-walk path
    \draw[myred, densely dashed, opacity=0.4] (I) -- (P0);
    \draw[myred, ->, >=stealth] (P0) -- (P1);
    \draw[myred, ->, >=stealth] (P1) -- (P2);
    \draw[myred, ->, >=stealth] (P2) -- (P3);
    \draw[myred, ->, >=stealth] (P3) -- (O);

    \node[draw=none, myblue, opacity=0.4, anchor=north,
          xshift=-0.8mm, yshift=0.5mm]
        at (P1) {\scalebox{0.4}{\comicneue PAF!}};
    \node[draw=none, myblue, opacity=0.4, anchor=north, yshift=0.5mm]
        at (P2) {\scalebox{0.4}{\comicneue PAF!}};
    \node[draw=none, myblue, opacity=0.4, anchor=north,
          xshift=0.8mm, yshift=0.5mm]
        at (P3) {\scalebox{0.4}{\comicneue PAF!}};

    % Centered legend
    \node[draw=none, anchor=center, inner sep=0pt, myred, opacity = 0.5]
        at (axis cs:-0.25,-0.95) {%
            \tikz[baseline=-0.5ex]{%
                \draw[myred, opacity=0.3, fill=myred, fill opacity=0.1]
                    (0,0) circle[radius=2mm];%
            }%
            \hspace{0.6mm}\scalebox{0.65}{$=$ phase function domain}%
        };
\end{axis}
\end{tikzpicture}}
        \\[-0.25em]
        (a) & (b)
    \end{tabular}
    \caption{
        (a) Heightfield versus (b) one-sided microflake medium. The characteristic transport is the same
        between the two. The volume preserves heightfield transport by requiring the phase function to 
        output directions within the spherical cap located above the incident direction. 
    }
    \label{fig:heightfield-medium-monotonicity}
\end{figure}

% =============================================================================================
\paragraph*{Transport from Specular Heightfield}
A specular heightfield produces a characteristic transport, which is illustrated in
Figure~\ref{fig:heightfield-medium-monotonicity}:
rays initially travel downwards, then each successive reflection increases their elevation until they escape.

% =============================================================================================
\paragraph*{One-Sided NDFs and Specular Reflection}
The characteristic transport described above results from two properties acting together.
First, every normal of a heightfield is strictly upwelling, which is represented by a one-sided NDF
\begin{equation}
    D(\boldsymbol{\omega}_m)=0
    \qquad\text{for }z_m\leq0.
    \label{eq:one-sided-ndf}
\end{equation}
Any microflake normal sampled from this distribution therefore satisfies $z_m>0$.
Second, each collision follows the specular reflection law
\begin{equation}
    \w_{j+1}
    =
    \w_j
    -
    2(\w_j\cdot\boldsymbol{\omega}_m)\boldsymbol{\omega}_m.
\end{equation}
The only way for a ray travelling along $\w_{j}$ to be reflected towards $\w_{j+1}$ at a lower elevation $z_{j+1}<z_j$ is
to encounter a downwelling normal, which is excluded by the one-sided NDF.
Hence, every reflection necessarily increases the ray elevation.

% =============================================================================================
\paragraph*{Path Sequences in the Microflake Volume}
The characteristic directional structure of heightfield transport depends only on the two properties identified above.
Both are directly available in our microflake volume: collisions are already specular, and their normal support is controlled by the NDF.
Reproducing this structure therefore requires nothing more than choosing a one-sided NDF.
Then, just as for a specular heightfield, any path within the microflake volume subject to $k-1$ collisions has an ordered 
direction sequence
\begin{equation}
    \Gamma_k=(\w_1,\w_2,\ldots,\w_k),
    \qquad
    z_1<z_2<\cdots<z_k,
    \label{eq:path-directions}
\end{equation}
with $z_1<0$ and $z_k>0$.
One-sided microflake volumes thus reproduce the ordered directional transport of a specular 
heightfield.

% =============================================================================================
\paragraph*{Spherical-Cap Support of the Phase Function}
The one-sided restriction also determines the admissible outgoing domain of the microflake phase function:
Since every reflection increases elevation, the phase function vanishes outside the spherical cap defined 
by the elevation of the incident direction
\begin{equation}
    \Psi(\w_j,\w_{j+1})=0
    \quad\text{for }z_{j+1}\leq z_j
    \qquad 
    \Rightarrow
    \qquad
    \w_{j+1}
    \in 
    \mathcal C(\w_j)
    =
    \left\{
        z_{j+1}>z_j
    \right\} \subset \mathcal S^2.
    \label{eq:cap}
\end{equation}
The spherical cap therefore defines the admissible support of the phase function, while the NDF determines how probability is distributed within it.

% =============================================================================================
% =============================================================================================
\subsection{Escape Probabilities and Scattering Orders}
\label{sec:path-shadowing}

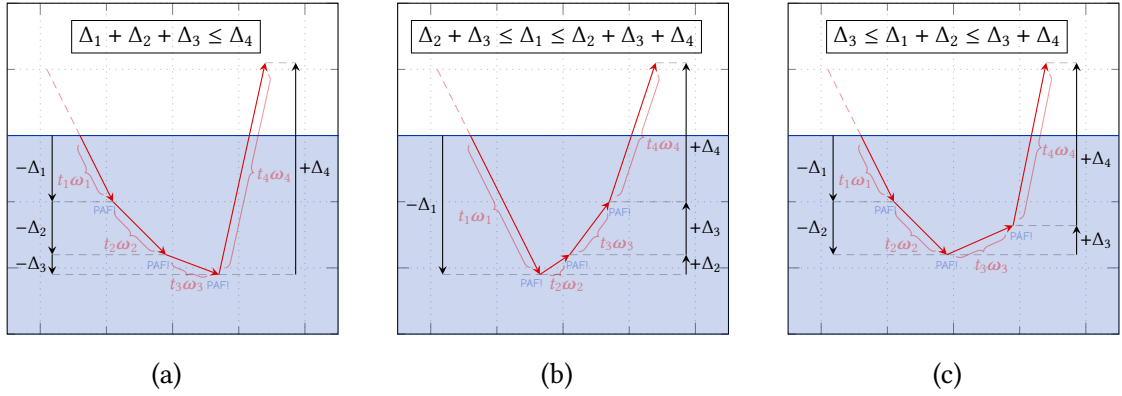
\begin{figure}[h]
    \centering
    \begin{tabular}{@{}ccc@{}}
        {\centering
\begin{tikzpicture}
\newcommand{\raysegment}[2]{%
    % segment
    \draw[myred, ->, >=stealth] (#1) -- (#2);

    % bracket
    \draw[myred, decorate, opacity=0.4,
          decoration={brace, raise=2pt,
                      pre=moveto, post=moveto,
                      pre length=4pt, post length=3pt}]
    (#2) -- (#1);
}

\begin{axis}[
    x = 1.75cm,
    y = 1.75cm,
    xmin = -2.25, xmax = 0.25,
    ymin = -1.5, ymax = 1,
    grid = both,
    axis line style = {color = black},
    minor grid style={dotted},
    major grid style={dotted},
    xticklabels = {},
    yticklabels = {},
    y label style = {rotate = -90},
    minor tick num = 1,
    xtick = {1,0,-1,-2},
    ytick = {1,0,-1,-2}
]
    % Semi-infinite medium
    \addplot[fill, color=myblue, opacity=0.2] coordinates
        {(-4,0) (4,0) (4,-4) (-4,-4)} -- cycle;
    \addplot[myblue, domain=-4:4] {0};
    
    \begin{scope}[shift={(axis direction cs: -0.2,0)}]
        % Path vertices: down, down, down, up
        \coordinate (I)  at (axis cs:-1.75,   0.5);
        \coordinate (P0) at (axis cs:-1.5, 0.0);
        \coordinate (P1) at (axis cs:-1.25, -0.5);
        \coordinate (P2) at (axis cs:-0.85,-0.9);
        \coordinate (P3) at (axis cs:-0.45,-1.05);
        \coordinate (P4) at (axis cs:-0.1,0.55);
    \end{scope}

        % Vertical-depth parameterization
        \coordinate (D1T) at (axis cs:-1.91,0.0);
        \coordinate (D1B) at (axis cs:-1.91,-0.5);
        \coordinate (D2T) at (axis cs:-1.91,-0.5);
        \coordinate (D2B) at (axis cs:-1.91,-0.9);
        \coordinate (D3T) at (axis cs:-1.91,-0.9);
        \coordinate (D3B) at (axis cs:-1.91,-1.05);
        \coordinate (D4B) at (axis cs:-0.07,-1.05);
        \coordinate (D4T) at (axis cs:-0.07,0.55);

        \draw[black, densely dashed, opacity=0.25] (D1B) -- (P1);
        \draw[black, ->, >=stealth]
            (D1T) -- node[left, xshift=0.7mm] {\scalebox{0.65}{$-\Delta_1$}} (D1B);
        \draw[black, densely dashed, opacity=0.25] (D2B) -- (P2);
        \draw[black, ->, >=stealth]
            (D2T) -- node[left, xshift=0.7mm] {\scalebox{0.65}{$-\Delta_2$}} (D2B);
        \draw[black, densely dashed, opacity=0.25] (D3B) -- (P3);
        \draw[black, ->, >=stealth]
            (D3T) -- node[left, xshift=0.7mm] {\scalebox{0.65}{$-\Delta_3$}} (D3B);
        \draw[black, densely dashed, opacity=0.25] (P3) -- (D4B);
        \draw[black, densely dashed, opacity=0.25] (P4) -- (D4T);
        \draw[black, ->, >=stealth]
            (D4B) -- node[left, xshift=6mm] {\scalebox{0.65}{$+\Delta_4$}} (D4T);

        % Triple-scattering path
        \draw[myred, densely dashed, opacity = 0.4] (I) -- (P0);
        \raysegment{P0}{P1}
        \raysegment{P1}{P2}
        \raysegment{P2}{P3}
        \raysegment{P3}{P4}

        \node[draw=none, myred, opacity=0.4, anchor=north west]
            at (axis cs:-1.95,-0.22) {\scalebox{0.65}{$t_1\w_1$}};
        \node[draw=none, myred, opacity=0.4, anchor=north west]
            at (axis cs:-1.60,-0.7) {\scalebox{0.65}{$t_2\w_2$}};
        \node[draw=none, myred, opacity=0.4, anchor=north west]
            at (axis cs:-1.1,-1.0) {\scalebox{0.65}{$t_3\w_3$}};
        \node[draw=none, myred, opacity=0.4, anchor=west]
            at (axis cs:-0.44,-0.30) {\scalebox{0.65}{$t_4\w_4$}};
        \node[draw=none, black, anchor=north]
            at (axis cs:-1.05, 0.95)
            {\scalebox{0.8}{$\boxed{\Delta_1+\Delta_2+\Delta_3\leq\Delta_4}$}};

        \node [opacity = 0.4, draw = none, myblue, anchor = north, yshift=+0.75mm, xshift=-1.1mm] at (P1) {\scalebox{0.4}{\comicneue PAF!}};
        \node [opacity = 0.4, draw = none, myblue, anchor = north, yshift=+0.5mm, xshift=-1.0mm] at (P2) {\scalebox{0.4}{\comicneue PAF!}};
        \node [opacity = 0.4, draw = none, myblue, anchor = north, yshift=+0.5mm, xshift=-0.1mm] at (P3) {\scalebox{0.4}{\comicneue PAF!}};

\end{axis}
\end{tikzpicture}}
        &
        {\centering
\begin{tikzpicture}
\newcommand{\raysegment}[2]{%
    % segment
    \draw[myred, ->, >=stealth] (#1) -- (#2);

    % bracket
    \draw[myred, decorate, opacity=0.4,
          decoration={brace, raise=2pt,
                      pre=moveto, post=moveto,
                      pre length=4pt, post length=3pt}]
    (#2) -- (#1);
}

\begin{axis}[
    x = 1.75cm,
    y = 1.75cm,
    xmin = -2.25, xmax = 0.25,
    ymin = -1.5, ymax = 1,
    grid = both,
    axis line style = {color = black},
    minor grid style={dotted},
    major grid style={dotted},
    xticklabels = {},
    yticklabels = {},
    y label style = {rotate = -90},
    minor tick num = 1,
    xtick = {1,0,-1,-2},
    ytick = {1,0,-1,-2}
]
    % Semi-infinite medium
    \addplot[fill, color=myblue, opacity=0.2] coordinates
        {(-4,0) (4,0) (4,-4) (-4,-4)} -- cycle;
    \addplot[myblue, domain=-4:4] {0};

    \begin{scope}[shift={(axis direction cs:-0.2,0)}]
        % Path vertices: down, up, up, up
        \coordinate (I)  at (axis cs:-1.75,0.5);
        \coordinate (P0) at (axis cs:-1.5,0.0);
        \coordinate (P1) at (axis cs:-0.975,-1.05);
        \coordinate (P2) at (axis cs:-0.75,-0.9);
        \coordinate (P3) at (axis cs:-0.45,-0.5);
        \coordinate (P4) at (axis cs:-0.1,0.55);
    \end{scope}

    % Vertical-depth parameterization
    \coordinate (D1T) at (axis cs:-1.91,0.0);
    \coordinate (D1B) at (axis cs:-1.91,-1.05);
    \coordinate (D2B) at (axis cs:-0.07,-1.05);
    \coordinate (D2T) at (axis cs:-0.07,-0.9);
    \coordinate (D3B) at (axis cs:-0.07,-0.9);
    \coordinate (D3T) at (axis cs:-0.07,-0.5);
    \coordinate (D4B) at (axis cs:-0.07,-0.5);
    \coordinate (D4T) at (axis cs:-0.07,0.55);

    \draw[black, densely dashed, opacity=0.25] (D1B) -- (P1);
    \draw[black, ->, >=stealth]
        (D1T) -- node[left, xshift=0.7mm] {\scalebox{0.65}{$-\Delta_1$}} (D1B);
    \draw[black, densely dashed, opacity=0.25] (P1) -- (D2B);
    \draw[black, densely dashed, opacity=0.25] (P2) -- (D2T);
    \draw[black, ->, >=stealth]
        (D2B) -- node[left, xshift=6mm] {\scalebox{0.65}{$+\Delta_2$}} (D2T);
    \draw[black, densely dashed, opacity=0.25] (P3) -- (D3T);
    \draw[black, ->, >=stealth]
        (D3B) -- node[left, xshift=6mm] {\scalebox{0.65}{$+\Delta_3$}} (D3T);
    \draw[black, densely dashed, opacity=0.25] (P4) -- (D4T);
    \draw[black, ->, >=stealth]
        (D4B) -- node[left, xshift=6mm, yshift=-2mm] {\scalebox{0.65}{$+\Delta_4$}} (D4T);

    % Triple-scattering path
    \draw[myred, densely dashed, opacity=0.4] (I) -- (P0);
    \raysegment{P0}{P1}
    \raysegment{P1}{P2}
    \raysegment{P2}{P3}
    \raysegment{P3}{P4}

    \node[draw=none, myred, opacity=0.4, anchor=north west]
        at (axis cs:-1.87,-0.49) {\scalebox{0.65}{$t_1\w_1$}};
    \node[draw=none, myred, opacity=0.4, anchor=north west]
        at (axis cs:-1.18,-1.00) {\scalebox{0.65}{$t_2\w_2$}};
    \node[draw=none, myred, opacity=0.4, anchor=north west]
        at (axis cs:-0.8,-0.69) {\scalebox{0.65}{$t_3\w_3$}};
    \node[draw=none, myred, opacity=0.4, anchor=west]
        at (axis cs:-0.45,-0.07) {\scalebox{0.65}{$t_4\w_4$}};
    \node[draw=none, black, anchor=north]
        at (axis cs:-1.05,0.95)
        {\scalebox{0.8}{$\boxed{\Delta_2+\Delta_3\leq\Delta_1\leq\Delta_2+\Delta_3+\Delta_4}$}};

    \node[opacity=0.4, draw=none, myblue, anchor=north, yshift=+0.75mm, xshift=-1.1mm]
        at (P1) {\scalebox{0.4}{\comicneue PAF!}};
    \node[opacity=0.4, draw=none, myblue, anchor=north, yshift=+0.5mm, xshift=1.0mm]
        at (P2) {\scalebox{0.4}{\comicneue PAF!}};
    \node[opacity=0.4, draw=none, myblue, anchor=north, yshift=+0.5mm, xshift=1.4mm]
        at (P3) {\scalebox{0.4}{\comicneue PAF!}};
\end{axis}
\end{tikzpicture}}
        &
        {\centering
\begin{tikzpicture}
\newcommand{\raysegment}[2]{%
    % segment
    \draw[myred, ->, >=stealth] (#1) -- (#2);

    % bracket
    \draw[myred, decorate, opacity=0.4,
          decoration={brace, raise=2pt,
                      pre=moveto, post=moveto,
                      pre length=4pt, post length=3pt}]
    (#2) -- (#1);
}

\begin{axis}[
    x = 1.75cm,
    y = 1.75cm,
    xmin = -2.25, xmax = 0.25,
    ymin = -1.5, ymax = 1,
    grid = both,
    axis line style = {color = black},
    minor grid style={dotted},
    major grid style={dotted},
    xticklabels = {},
    yticklabels = {},
    y label style = {rotate = -90},
    minor tick num = 1,
    xtick = {1,0,-1,-2},
    ytick = {1,0,-1,-2}
]
    % Semi-infinite medium
    \addplot[fill, color=myblue, opacity=0.2] coordinates
        {(-4,0) (4,0) (4,-4) (-4,-4)} -- cycle;
    \addplot[myblue, domain=-4:4] {0};
    
    \begin{scope}[shift={(axis direction cs: -0.2,0)}]
        % Path vertices: down, down, down, up
        \coordinate (I)  at (axis cs:-1.75,   0.5);
        \coordinate (P0) at (axis cs:-1.5, 0.0);
        \coordinate (P1) at (axis cs:-1.25, -0.5);
        \coordinate (P2) at (axis cs:-0.85,-0.9);
        \coordinate (P3) at (axis cs:-0.35,-0.68);
        \coordinate (P4) at (axis cs:-0.1,0.55);
    \end{scope}

        % Vertical-depth parameterization
        \coordinate (D1T) at (axis cs:-1.91,0.0);
        \coordinate (D1B) at (axis cs:-1.91,-0.5);
        \coordinate (D2T) at (axis cs:-1.91,-0.5);
        \coordinate (D2B) at (axis cs:-1.91,-0.9);
        \coordinate (D3T) at (axis cs:-0.07,-0.9);
        \coordinate (D3B) at (axis cs:-0.07,-0.68);
        \coordinate (D4B) at (axis cs:-0.07,-0.68);
        \coordinate (D4T) at (axis cs:-0.07,0.55);

        \draw[black, densely dashed, opacity=0.25] (D1B) -- (P1);
        \draw[black, ->, >=stealth]
            (D1T) -- node[left, xshift=0.7mm] {\scalebox{0.65}{$-\Delta_1$}} (D1B);
        \draw[black, densely dashed, opacity=0.25] (D2B) -- (P2);
        \draw[black, ->, >=stealth]
            (D2T) -- node[left, xshift=0.7mm] {\scalebox{0.65}{$-\Delta_2$}} (D2B);
        \draw[black, densely dashed, opacity=0.25] (D3B) -- (P3);
        \draw[black, ->, >=stealth]
            (D3T) -- node[left, xshift=6mm] {\scalebox{0.65}{$+\Delta_3$}} (D3B);
        \draw[black, densely dashed, opacity=0.25] (P2) -- (D3T);
        \draw[black, densely dashed, opacity=0.25] (P4) -- (D4T);
        \draw[black, ->, >=stealth]
            (D4B) -- node[left, xshift=6mm, yshift=-2mm] {\scalebox{0.65}{$+\Delta_4$}} (D4T);

        % Triple-scattering path
        \draw[myred, densely dashed, opacity = 0.4] (I) -- (P0);
        \raysegment{P0}{P1}
        \raysegment{P1}{P2}
        \raysegment{P2}{P3}
        \raysegment{P3}{P4}

        \node[draw=none, myred, opacity=0.4, anchor=north west]
            at (axis cs:-1.95,-0.22) {\scalebox{0.65}{$t_1\w_1$}};
        \node[draw=none, myred, opacity=0.4, anchor=north west]
            at (axis cs:-1.60,-0.7) {\scalebox{0.65}{$t_2\w_2$}};
        \node[draw=none, myred, opacity=0.4, anchor=north west]
            at (axis cs:-0.92,-0.86) {\scalebox{0.65}{$t_3\w_3$}};
        \node[draw=none, myred, opacity=0.4, anchor=west]
            at (axis cs:-0.41,-0.12) {\scalebox{0.65}{$t_4\w_4$}};
        \node[draw=none, black, anchor=north]
            at (axis cs:-1.05, 0.95)
            {\scalebox{0.8}{$\boxed{\Delta_3 \leq \Delta_1+\Delta_2 \leq \Delta_3+\Delta_4}$}};

        \node [opacity = 0.4, draw = none, myblue, anchor = north, yshift=+0.75mm, xshift=-1.1mm] at (P1) {\scalebox{0.4}{\comicneue PAF!}};
        \node [opacity = 0.4, draw = none, myblue, anchor = north, yshift=+0.5mm, xshift=0.0mm] at (P2) {\scalebox{0.4}{\comicneue PAF!}};
        \node [opacity = 0.4, draw = none, myblue, anchor = north, yshift=+0.5mm, xshift=1.1mm] at (P3) {\scalebox{0.4}{\comicneue PAF!}};

\end{axis}
\end{tikzpicture}}
        \\[-0.25em]
        (a) & (b) & (c)
    \end{tabular}
    \caption{ \label{fig:triple-scattering} All three possible path configurations for escaping the medium after 3 collisions. }
\end{figure}

% =============================================================================================
\paragraph{Escape Probability After Three Collisions}
The characteristic motion of heightfield-like transport makes it possible to enumerate the 
cases that occur to determine the exit probability of, e.g., triple scattering.
As illustrated in Figure~\ref{fig:triple-scattering}, only three regimes are possible:
\begin{equation}
\begin{array}{lll}
z_2<z_3<0, &\quad&
z_2<0<z_3,
\qquad
0<z_2<z_3.
\end{array}
\label{eq:three-path-regimes}
\end{equation}
The two pure regimes factor into simple products,
\begin{equation}
P_3(\w_1, \cdots, \w_4)\big|_{z_3<0}
=\frac{\kappa_1}{\kappa_1+\kappa_4}
 \frac{\kappa_2}{\kappa_2+\kappa_4}
 \frac{\kappa_3}{\kappa_3+\kappa_4}
,
\label{eq:P3-pure1}
\end{equation}
and
\begin{equation}
P_3(\w_1, \cdots, \w_4)\big|_{z_2>0}
=\frac{\kappa_1}{\kappa_1+\kappa_4}
 \frac{\kappa_2}{\kappa_1+\kappa_2}
 \frac{\kappa_3}{\kappa_1+\kappa_3}.
\label{eq:P3-pure}
\end{equation}
Finally, the mixed regime gives
\begin{align}
P_3(\w_1, \cdots, \w_4)\big|_{z_2<0<z_3}
&=\int_0^\infty p_1(\Delta_1) \,
  \int_0^\infty p_2(\Delta_2) \,
  \int_0^{\Delta_1+\Delta_2}p_3(\Delta_3) \,
  \int_{\Delta_1+\Delta_2-\Delta_3}^\infty p_4(\Delta_4) \,
  \dd \Delta_4 \, \dd \Delta_3\, \dd \Delta_2 \, \dd \Delta_1
  \notag
  \\
&=
\frac{\kappa_1\kappa_2\kappa_3}
{(\kappa_1+\kappa_4)(\kappa_2+\kappa_3)}
\left(
\frac{1}{\kappa_2+\kappa_4}
+\frac{1}{\kappa_1+\kappa_3}
\right).
\label{eq:P3-mixed}
\end{align}

% =============================================================================================
\paragraph{General Escape Probability}
Noticing that Equation~\eqref{eq:P3-mixed} can be re-written in the same spirit as 
Equation~\eqref{eq:P2} 
\begin{equation}
P_3(\w_1,\w_2,\w_3,\w_4)
=
\frac{
\ind_{z_2<0}\,\kappa_1 \,P_2(\w_2,\w_3,\w_4)
+
\ind_{z_3>0}\,\kappa_3 \,P_2(\w_1,\w_2,\w_3)
}{\kappa_1+\kappa_4},
\label{eq:P3-recursive}
\end{equation}
we obtain the escape probability for any number of collisions
\begin{equation}
P_k(\w_1,\ldots,\w_{k+1})
=
\begin{cases}
\displaystyle \frac{\kappa_1}{\kappa_1+\kappa_2}, & k=1,\\[8pt]
\displaystyle \frac{
\ind_{z_2<0}\,\kappa_1 \, P_{k-1}(\w_2,\ldots,\w_{k+1})
+
\ind_{z_k>0}\,\kappa_k \, P_{k-1}(\w_1,\ldots,\w_k)
}{\kappa_1+\kappa_{k+1}}, & \text{otherwise}.
\end{cases}
\label{eq:path-shadowing-recursion}
\end{equation}
Equation~\eqref{eq:path-shadowing-recursion} is a special case of the semi-infinite exit 
probability of Bitterli and d'Eon~\cite[Eq.~(22)]{bitterli2022positionfree}. It is also closely related 
to the path recurrence of Cui~et~al.~\cite[Eq.~(14)]{cui2023invariance}, which is derived from the invariance 
principle (whereas ours follows directly from the geometry of the transport paths).

% =============================================================================================
\paragraph{General Exit-Direction Density}
For a fixed incident direction, the order-$k$ contribution to the exit-direction density is obtained by marginalizing all intermediate directions:
\begin{equation}
\mathcal{R}_k(\w_1, \w_{k+1})
=\int_{z_1<z_2<\cdots<z_{k+1}}
P_k(\w_1,\ldots,\w_{k+1})
\prod_{j=1}^{k}\Psi(\w_j, \w_{j+1})
\prod_{j=2}^{k}\dd\w_j.
\label{eq:path-order-response}
\end{equation}
For $k=1$ and $k=2$, this expression recovers Equations~\eqref{eq:generic-single} and~\eqref{eq:generic-double}; together with the escape probability derived above, it also defines the third-order density.
The complete outgoing-direction density is the physical scattering-order series
\begin{equation}
    \boxed{\mathcal R(\w_1, \w_o)=\sum_{k=1}^{\infty}\mathcal{R}_k(\w_1, \w_o).}
    \label{eq:path-series}
\end{equation}
Solving Equation~\eqref{eq:path-series} is now a matter of finding an NDF 
for which the intermediate directions in Eq.~\eqref{eq:path-order-response} can be 
marginalized analytically and the resulting order series can be summed in closed form.

% =============================================================================================
% =============================================================================================
\subsection{Optional -- Uniform-Cap Scattering Orders}
\label{sec:heightfield:optional}

The derivation in this subsection is not required for the rest of the paper.
Its goal is to illustrate how the framework developed so far efficiently recovers known results for heightfield transport.

\paragraph*{Uniform-Cap Scattering}
We apply it to the simplest one-sided microflake NDF, obtained by using a constant distribution over the upwelling hemisphere:
\begin{equation}
    D(\w_m)
    =
    \frac{\ind_{z_m>0}}{\pi}
    \quad\Rightarrow\quad
    \sigma(\w_j)
    =
    \frac{1-z_j}{2},
    \qquad
    \kappa(\w_j)
    =
    \frac{1-z_j}{2|z_j|}.
    \label{eq:ggx1-properties}
\end{equation}
Inserting Equation~\eqref{eq:ggx1-properties} into the microflake phase function of Equation~\eqref{eq_phase} gives
\begin{equation}
    \Psi(\w_j,\w_{j+1})
    =
    \frac{\ind_{z_{j+1}>z_j}}{2\pi(1-z_j)}.
    \label{eq:ggx1-phase}
\end{equation}
This phase function is uniform over the spherical cap $\mathcal C(\w_j)$ as defined in Equation~\eqref{eq:cap} 
and leads to the configuration of a GGX microfacet distribution with roughness $\alpha=1$~\cite{dupuy2023caps}.
We now derive the first two scattering orders of the corresponding BRDF.

% =============================================================================================
\paragraph*{Single-scattering BRDF}
Inserting Equations~\eqref{eq:ggx1-properties} and~\eqref{eq:ggx1-phase} into Equation~\eqref{eq:generic-single} gives
\begin{align}
    \mathcal R_1(\w_1,\w_o)
    &=
    \frac{\kappa_1}{\kappa_1+\kappa_o}
    \frac{1}{2\pi(1-z_1)}
    \notag\\
    &=
    \frac{1}{2\pi}
    \frac{z_o}{z_o-z_1}.
    \label{eq:ggx1-R1}
\end{align}
We then retrieve the single-scattering BRDF from Equation~\eqref{eq:brdf-link}:
\begin{equation}
    f_{r_1}(\w_i,\w_o)
    =
    \frac{\mathcal R_1(-\w_i,\w_o)}{z_o}
    =
    \frac{1}{2\pi}
    \frac{1}{z_i+z_o}.
    \label{eq:ggx1-f1}
\end{equation}
This single-scattering BRDF is exactly twice the Chandrasekhar single-scattering term of Equation~\eqref{eq:chandrasekhar-f1}.
\vspace{1cm}
% =============================================================================================
\paragraph*{Double-scattering BRDF}
Inserting Equations~\eqref{eq:P2}, \eqref{eq:ggx1-properties}, and~\eqref{eq:ggx1-phase} into Equation~\eqref{eq:generic-double} gives
\begin{align}
    \mathcal R_2(\w_1,\w_o)
    &=
    \mathcal R_1(\w_1,\w_o)
    \left[
        z_o
        \int_{z_1}^{0}
        \frac{1}{z_o-z_2}
        \,\dd z_2
        -
        z_1
        \int_{0}^{z_o}
        \frac{1}{z_2-z_1}
        \,\dd z_2
    \right]
    \notag\\
    &=
    \mathcal R_1(\w_1,\w_o)
    \left[
        z_o
        \log\left(\frac{z_o-z_1}{z_o}\right)
        -
        z_1
        \log\left(\frac{z_o-z_1}{-z_1}\right)
    \right].
    \label{eq:ggx1-R2}
\end{align}
We then retrieve the double-scattering BRDF from Equation~\eqref{eq:brdf-link}:
\begin{align}
    f_{r_2}(\w_i,\w_o)
    &=
    \frac{\mathcal R_2(-\w_i,\w_o)}{z_o}
    \notag\\
    &=
    f_{r_1}(\w_i,\w_o)
    \left[
        z_i
        \log\left(\frac{z_i+z_o}{z_i}\right)
        +
        z_o
        \log\left(\frac{z_i+z_o}{z_o}\right)
    \right]
    \notag\\
    &=
    \frac{1}{2 \pi}
    \frac{
        z_i\log\left(\frac{z_i+z_o}{z_i}\right)
        +
        z_o\log\left(\frac{z_i+z_o}{z_o}\right)
    }{
        z_i+z_o
    }.
    \label{eq:ggx1-f2}
\end{align}
Equation~\eqref{eq:ggx1-f1} is the known single-scattering GGX BRDF at unit roughness, whereas Equation~\eqref{eq:ggx1-f2} provides, to our knowledge, a new closed-form expression for its double-scattering order.
This new result follows directly from the framework developed above.
\vspace{1cm}
% =============================================================================================
\paragraph*{Higher-Order Densities}
Unfortunately, subsequent scattering orders become increasingly complicated despite the simplicity of the phase function.
This is also true for the Chandrasekhar scattering laws derived in Section~\ref{sec:generic:optional}.
Consequently, the uniform-cap NDF does not yield an elementary closed form for the complete scattering-order series.
In the next section, we derive another one-sided NDF for which all intermediate directions can be marginalized and the complete series can be summed in closed form.

% =============================================================================================
% =============================================================================================
% =============================================================================================
\section{Elementary Solution Using a Quadratic NDF}
\label{sec:quadratic}

Here we introduce an NDF that leads to an elementary expression for the BRDF. 

% =============================================================================================
% =============================================================================================
\subsection{The Quadratic NDF}

% =============================================================================================
\paragraph*{Definition}
We specialize the transport model to the parameter-free quadratic NDF:
\begin{equation}
    D(\w_m)
    =
    \frac{2}{\pi}z_m^2\ind_{z_m>0}
    \quad\Rightarrow\quad
    \sigma(\w_j)
    =
    \frac{(1-z_j)^2}{4},
    \qquad
    \kappa(\w_j)
    =
    \frac{(1-z_j)^2}{4|z_j|}.
    \label{eq:D2}
\end{equation}
Inserting Equation~\eqref{eq:D2} into the microflake phase function of Equation~\eqref{eq_phase} gives
\begin{align}
    \Psi(\w_j,\w_{j+1})
    &=
    \frac{1}{4\sigma(\w_j)}
    \frac{2}{\pi}
    \left(
    \frac{z_{j+1}-z_j}{\|\w_{j+1}-\w_j\|}
    \right)^2
    \notag\\
    &=
    \frac{1}{4\sigma(\w_j)}
    \frac{2}{\pi}
    \frac{(z_{j+1}-z_j)^2}
    {2(1-\w_j\cdot\w_{j+1})}
    \notag\\
    &=
    \frac{(z_{j+1}-z_j)^2}
    {\pi(1-z_j)^2(1-\w_j\cdot\w_{j+1})}.
    \label{eq:quad-phase-compact}
\end{align}

% =============================================================================================
\paragraph*{Link to the Poisson Kernel}
Equation~\eqref{eq:quad-phase-compact} can be factorized as
\begin{equation}
    \Psi(\w_j,\w_{j+1})
    =h(z_{j+1}\mid z_j) \, \Ppois_{r_j}(\phi_{j+1}-\phi_j),
    \label{eq:transport-phase-factorization}
\end{equation}
where $h$ is the univariate density over the elevation $z_{j+1}\in[z_j,1]$ of the outgoing direction
\begin{equation}
    h(z_{j+1}\mid z_j)=\frac{2(z_{j+1}-z_j)}{(1-z_j)^2},
    \label{eq:transport-elevation-density}
\end{equation}
and $\Ppois_{r_j}$ is the circular Poisson kernel whose parameter is obtained from the elevations through
\begin{equation}
\begin{aligned}
    \Ppois_{r_j}(\phi_{j+1}-\phi_j)
    &=\frac{1-r_j^2}{2\pi\left(1-2r_j\cos(\phi_{j+1}-\phi_j)+r_j^2\right)}
    =\frac{z_{j+1}-z_j}{2\pi(1-\w_j\cdot\w_{j+1})},
    \\
    r_j&=\frac{q(z_{j+1})}{q(z_j)},
    \\
    q(z)&=\sqrt{\frac{1-z}{1+z}}.
\end{aligned}
    \label{eq:poisson}
\end{equation}
By construction, $r_j\in[0,1]$ for every valid transition $z_{j+1}>z_j$.
We now exploit this factorization to solve the exit-direction density of Equation~\eqref{eq:path-order-response} and its scattering-order series in Equation~\eqref{eq:path-series}.

% =============================================================================================
% =============================================================================================
\subsection{Azimuthal Marginalization of an Arbitrary Path}

\paragraph*{Stability under Circular Convolution}
The Poisson kernel is stable under circular convolution:
\begin{equation}
    \Ppois_r*\Ppois_s=\Ppois_{rs},
    \label{eq:poisson-semigroup}
\end{equation}
This property allows us to marginalize the azimuthal domain of Equation~\eqref{eq:path-order-response} analytically.

\paragraph*{Azimuthal Marginalization}
Since the NDF is radially symmetric, the escape probability $P_k$ depends only on the elevations of the path directions.
This is also true for the elevation density $h$.
Letting $\w_{k+1}=\w_o$, we insert the factorization of Equation~\eqref{eq:transport-phase-factorization} into Equation~\eqref{eq:path-order-response} and obtain
\begin{align}
    \mathcal R_k(\w_1,\w_o)
    &=
    \int_{z_1<z_2<\cdots<z_k<z_o}
    P_k(z_1,\ldots,z_k,z_o)
    \prod_{j=1}^{k}h(z_{j+1}\mid z_j)
    \notag\\
    &\quad\times
    \underbrace{
    \left[
    \int_0^{2\pi}\!\cdots\!\int_0^{2\pi}
    \prod_{j=1}^{k}
    \Ppois_{r_j}(\phi_{j+1}-\phi_j)
    \dd\phi_2\cdots\dd\phi_k
    \right]
    }_{\text{circular convolutions}}
    \prod_{j=2}^{k}\dd z_j,
    \label{eq:azimuth-path-integral}
\end{align}
where $z_{k+1}=z_o$ and $\phi_{k+1}=\phi_o$.
Thanks to the stability property of the Poisson kernel, the sequence of circular convolutions reduces to
\begin{equation}
    \left(
    \Ppois_{r_1}*\cdots*\Ppois_{r_k}
    \right)(\phi_o-\phi_1)
    =
    \Ppois_{\prod_{j=1}^{k}r_j}(\phi_o-\phi_1).
    \label{eq:azimuth-path-convolution}
\end{equation}

\paragraph*{Telescoping along the Path}
Equation~\eqref{eq:azimuth-path-convolution} can be simplified further since
\begin{equation}
    \prod_{j=1}^{k}r_j
    =
    \prod_{j=1}^{k}
    \frac{q(z_{j+1})}{q(z_j)}
    =
    \frac{q(z_o)}{q(z_1)} =: \eta.
    \label{eq:poisson-telescope}
\end{equation}
The sequence of circular convolutions thus only depends on the elevations of the incident and outgoing directions.
It follows that the Poisson kernel can be taken outside the remaining elevation integral.
\begin{align}
    \mathcal R_k(\w_1,\w_o)
    &=
    z_o \, F_k(z_1,z_o)
    \,\Ppois_\eta(\phi_o-\phi_1)
    \notag\\
    &=
    z_o\,F_k(z_1,z_o)\,
    \frac{z_o-z_1}{2\pi(1-\w_1\cdot\w_o)},
    \label{eq:order-poisson-factorization}
\end{align}
where
\begin{equation}
    z_o\,F_k(z_1,z_o)
    =
    \int_{z_1<z_2<\cdots<z_k<z_o}
    P_k(z_1,\ldots,z_k,z_o)
    \prod_{j=1}^{k}h(z_{j+1}\mid z_j)
    \prod_{j=2}^{k}\dd z_j.
    \label{eq:elevation-amplitude}
\end{equation}

% =============================================================================================
% =============================================================================================
\subsection{One-Dimensional Scattering-Order Series}

\paragraph*{Single-Scattering Amplitude}
After the azimuthal marginalization, it remains only to determine the elevation amplitudes defined in Equation~\eqref{eq:elevation-amplitude}.
We parameterize their endpoint dependence by
\begin{equation}
    p:=-z_1z_o\in[0,1].
\end{equation}
For $k=1$, Equation~\eqref{eq:elevation-amplitude} gives
\begin{align}
    z_oF_1(z_1,z_o)
    &=
    P_1(z_1,z_o)h(z_o\mid z_1)
    \notag\\
    &=
    \frac{\kappa_1}{\kappa_1+\kappa_o}
    \frac{2(z_o-z_1)}{(1-z_1)^2}
    \notag\\
    &=
    \frac{2z_o}{1-z_1z_o},
    \notag\\
    F_1(z_1,z_o)
    &=
    S_1(p)
    =
    \frac{2}{1+p}.
    \label{eq:quadratic-S1}
\end{align}
The factor $z_o$ therefore cancels, leaving an amplitude that depends on the endpoint elevations only through $p$.

\paragraph*{Scattering-Order Recurrence}
We now show that this dependence is preserved at every scattering order.
Assume for some $k\geq1$ that
\begin{equation}
    F_k(z_1,z_o)=S_k(-z_1z_o).
\end{equation}
Substituting the escape-probability recurrence of Equation~\eqref{eq:path-shadowing-recursion} into the elevation integral of Equation~\eqref{eq:elevation-amplitude} gives the amplitude $F_{k+1}$.
For the quadratic NDF, the factors involving the newly integrated elevation $z$ simplify to
\begin{equation}
    \kappa_1+\kappa_o
    =
    \frac{(z_o-z_1)(1-z_1z_o)}{-4z_1z_o},
    \qquad
    \kappa_1h(z\mid z_1)
    =
    \frac{z-z_1}{-2z_1},
    \qquad
    \kappa(z)h(z_o\mid z)
    =
    \frac{z_o-z}{2z}.
    \label{eq:quadratic-recursion-factors}
\end{equation}
Using $t=-zz_o$ on the downwelling branch and $t=-z_1z$ on the upwelling branch maps both contributions to the interval $0<t<p$.
Combining them gives the scalar recurrence
\begin{align}
    S_{k+1}(p)
    &=
    \frac{2}{p(1+p)}
    \int_0^p
    (p-t)S_k(t)\dd t
    \notag\\
    &=
    \frac{2p}{1+p}
    \int_0^1
    (1-u)S_k(pu)\dd u
    =:
    (\mathcal T S_k)(p).
    \label{eq:scattering-order-operator}
\end{align}
Since the result depends only on $p$, the induction closes and every scattering order has the form
\begin{equation}
    \mathcal R_k(\w_1,\w_o)
    =
    z_oS_k(p)
    \Ppois_\eta(\phi_o-\phi_1).
\end{equation}

\paragraph*{Telescoping Scattering-Order Series}
The source term $S_1$ can itself be expressed using the recurrence operator.
Indeed, applying $\mathcal T$ to the constant function $2$ gives
\begin{equation}
    (\mathcal T2)(p)
    =
    \frac{2p}{1+p},
    \qquad
    S_1(p)
    =
    2-(\mathcal T2)(p),
\end{equation}
or equivalently
\begin{equation}
    S_1=(I-\mathcal T)2.
\end{equation}
Since $S_k=\mathcal T^{k-1}S_1$, each order can be written as the difference
\begin{equation}
    S_k
    =
    \mathcal T^{k-1}2-\mathcal T^k2,
    \qquad
    \sum_{k=1}^{K}S_k(p)
    =
    2-\mathcal T^K2(p).
    \label{eq:partial-sum}
\end{equation}
For any bounded function $f$ and $0\leq p\leq1$, the operator satisfies
\begin{equation}
    |(\mathcal T f)(p)|
    \leq
    \frac{p}{1+p}\|f\|_\infty
    \leq
    \frac{1}{2}\|f\|_\infty.
\end{equation}
Since $\mathcal T$ preserves non-negativity, the truncation error is bounded by
\begin{equation}
    0
    \leq
    2-\sum_{k=1}^{K}S_k(p)
    =
    \mathcal T^K2(p)
    \leq
    2^{1-K}.
    \label{eq:truncation-bound}
\end{equation}
The scattering-order series therefore converges uniformly to
\begin{equation}
    S(p)
    :=
    \sum_{k=1}^{\infty}S_k(p)
    =
    2.
    \label{eq:S2}
\end{equation}

\paragraph*{Complete BRDF}
Combining this result with Equations~\eqref{eq:path-series} and~\eqref{eq:order-poisson-factorization} gives the complete transport response
\begin{align}
    \mathcal R(\w_1,\w_o)
    &=
    z_o
    \left[
    \sum_{k=1}^{\infty}S_k(p)
    \right]
    \Ppois_\eta(\phi_o-\phi_1)
    \notag\\
    &=
    2z_o
    \Ppois_\eta(\phi_o-\phi_1)
    \notag\\
    &=
    2z_o
    \frac{z_o-z_1}{2\pi(1-\w_1\cdot\w_o)}.
    \label{eq:render-poisson-response}
\end{align}
Returning to the conventional outward-pointing incident direction with $\w_i=-\w_1$ and $z_i=-z_1$, we retrieve the BRDF from Equation~\eqref{eq:brdf-link}:
\begin{align}
    f_r(\w_i,\w_o)
    &=
    \frac{\mathcal R(-\w_i,\w_o)}{z_o}
    \notag\\
    &=
    2\Ppois_\eta(\phi_o-\phi_i-\pi)
    \notag\\
    &=
    2\frac{z_i+z_o}{2\pi(1+\w_i\cdot\w_o)}
    \notag\\
    &=
    \frac{z_i+z_o}
    {\pi(1+\w_i\cdot\w_o)}.
    \label{eq:render-compact}
\end{align}

% =============================================================================================
% =============================================================================================
% =============================================================================================
\section{Importance Sampling}
\label{sec:sampling}

We exploit the connection established in Equation~\eqref{eq:render-compact} between the BRDF and the Poisson kernel to derive a direct sampling routine.

% =============================================================================================
% =============================================================================================
\subsection{Sampling the BRDF}

% =============================================================================================
\paragraph*{Poisson-Kernel Sampling}
Let $v=e^{\mathrm i\phi}$ be a point on the unit circle and let
$\eta=x+\mathrm i y$ be a complex parameter such that $|\eta|<1$.
The circular Poisson kernel with parameter $\eta$ is
\begin{equation}
    \Ppois(v;\eta)
    :=
    \frac{1-|\eta|^2}{2\pi|v-\eta|^2}.
    \label{eq:complex-poisson-kernel}
\end{equation}
Its magnitude $|\eta|$ controls the concentration and its argument $\arg\eta$ the orientation.
McCullagh~\cite{mccullagh1996mobius} samples this kernel geometrically by using $\eta$ as a pivot inside the disk.
Given a uniform point $c=e^{\mathrm i\theta}$ on the unit circle, the line passing through $c$ and $\eta$ intersects the circle again at
\begin{equation}
    v
    =
    \frac{\eta-c}{1-\overline\eta\,c}.
    \label{eq:chord-transform}
\end{equation}
This transformation maps the unit circle onto itself and its angular Jacobian is
\begin{equation}
    \frac{1}{2\pi}
    \left|\frac{\partial\theta}{\partial\phi}\right|
    =
    \frac{1-|\eta|^2}{2\pi|v-\eta|^2}
    =
    \Ppois(v;\eta).
    \label{eq:chord-jacobian}
\end{equation}
Thus, the transformed point $v$ follows the Poisson kernel of parameter $\eta$.

\begin{figure}
    \centering
\begin{tikzpicture}
    \def\diskradius{2.0}

    \newcommand{\drawcontourdisk}[3]{%
        \fill[#2, opacity=#3]
            (0,0) circle[radius={#1*\diskradius cm}];
        \draw[mygreylight]
            (0,0) circle[radius={#1*\diskradius cm}];
    }

    \newcommand{\drawdisk}[3]{%
        \fill[#2, opacity=#3]
            (0,0) circle[radius={#1*\diskradius cm}];
    }

    \newcommand{\drawcontourmapping}[2]{%
        % Incident direction: x_i=-sin(75 degrees), z_i=cos(75 degrees).
        \pgfmathsetmacro{\qi}{tan(37.5)}
        \pgfmathsetmacro{\qo}{#1/(1+sqrt(1-#1*#1))}
        \pgfmathsetmacro{\pivot}{\qi*\qo}
        \pgfmathsetmacro{\pivotx}{#1*\diskradius*\pivot}

        \foreach \sample in {0,...,6} {
            \pgfmathsetmacro{\phi}{#2 + 360*\sample/7}
            \pgfmathsetmacro{\costheta}{cos(\phi)}
            \pgfmathsetmacro{\sintheta}{sin(\phi)}
            \pgfmathsetmacro{\chordparameter}{%
                2*(1-\pivot*\costheta)
                /(1+\pivot*\pivot-2*\pivot*\costheta)%
            }
            \pgfmathsetmacro{\originalx}{#1*\diskradius*\costheta}
            \pgfmathsetmacro{\originaly}{#1*\diskradius*\sintheta}
            \pgfmathsetmacro{\transformedx}{%
                #1*\diskradius*(\costheta
                +\chordparameter*(\pivot-\costheta))%
            }
            \pgfmathsetmacro{\transformedy}{%
                #1*\diskradius*(\sintheta
                -\chordparameter*\sintheta)%
            }

            \draw[mygrey, opacity=0.1]
                (\originalx,\originaly)
                -- (\transformedx,\transformedy);
            \node[draw=mygrey, circle, fill=mygreylighter,
                  inner sep=0.675pt]
                at (\originalx,\originaly) {};
            \node[draw=myorange, circle, fill=myorangelight,
                  inner sep=0.675pt]
                at (\transformedx,\transformedy) {};
        }

        \node[draw=myred, circle, fill=myredlighter, inner sep=1.0pt]
            at (\pivotx,0) {};
    }

    \newcommand{\drawtransformedcontour}[3]{%
        \begin{scope}[shift={(0,#3cm)}]
            \begin{scope}[yscale=0.45]
                \drawcontourdisk{#1}{mygreylighter}{1.0}
                \drawcontourmapping{#1}{#2}
            \end{scope}
        \end{scope}
    }

    \newcommand{\drawshadow}[3]{%
        \begin{scope}[shift={(0,#3cm)}]
            \begin{scope}[yscale=0.45]
                \drawdisk{#1}{mygrey}{0.1}
            \end{scope}
        \end{scope}
    }

    \newcommand{\drawfinalsamples}[2]{%
        \pgfmathsetmacro{\qi}{tan(37.5)}
        \pgfmathsetmacro{\qo}{#1/(1+sqrt(1-#1*#1))}
        \pgfmathsetmacro{\pivot}{\qi*\qo}

        \foreach \sample in {0,...,6} {
            \pgfmathsetmacro{\phi}{#2 + 360*\sample/7}
            \pgfmathsetmacro{\costheta}{cos(\phi)}
            \pgfmathsetmacro{\sintheta}{sin(\phi)}
            \pgfmathsetmacro{\chordparameter}{%
                2*(1-\pivot*\costheta)
                /(1+\pivot*\pivot-2*\pivot*\costheta)%
            }
            \pgfmathsetmacro{\transformedx}{%
                #1*\diskradius*(\costheta
                +\chordparameter*(\pivot-\costheta))%
            }
            \pgfmathsetmacro{\transformedy}{%
                #1*\diskradius*(\sintheta
                -\chordparameter*\sintheta)%
            }

            \node[draw=myorange, circle, fill=myorangelight,
                  inner sep=0.675pt]
                at (\transformedx,\transformedy) {};
        }
    }

    % Hemisphere above its orthographic disk projection
    \draw[mygreylight, fill = mygreylighter]
        (-\diskradius cm,0)
        arc[start angle=180, end angle=0, radius=\diskradius cm];

    \begin{scope}[yscale=0.45]
        % Oblique orthographic view of the disk and its iso-radius contours
        \fill[mygreylighter] (0,0) circle[radius=\diskradius cm];
        \foreach \rho in {0.333333,0.666667,1.00} {
            \draw[mygreylight]
                (0,0) circle[radius={\rho*\diskradius cm}];
        }

        % Seven uniform azimuthal samples with a phase shift on each contour
        \foreach \rho/\offset in {
            0.333333/0,
            0.666667/17.143,
            1.00/34.286
        } {
            \foreach \sample in {0,...,6} {
                \pgfmathsetmacro{\phi}{\offset + 360*\sample/7}
                \node[draw=mygrey, circle, fill=mygreylighter, inner sep=0.675pt]
                    at (\phi:{\rho*\diskradius cm}) {};
            }
        }
    \end{scope}

    % Incident direction: w_i=(-sin(75 degrees),0,cos(75 degrees))
    \draw[myred, ->, >=stealth]
        (0,0) -- (165:\diskradius cm)
        node[pos=0.62, anchor=north west]
            {\scalebox{0.65}{$\w_i$}};
    \node[draw=none, anchor=center, inner sep=0pt]
        at (0,-1.20cm) {%
            \tikz[baseline=-0.5ex]{%
                \node[draw=mygrey, circle, fill=mygreylighter,
                      inner sep=0.675pt] {};%
            }%
            \hspace{0.5mm}\scalebox{0.7}{$\sim\mathcal{D}$}%
        };

    % Exploded per-radius chord transformations for theta_i = 75 degrees
    \begin{scope}[xshift=5cm, yshift = 0cm]
        \draw[mygreylight, fill = mygreylighter]
            (-\diskradius cm,0)
            arc[start angle=180, end angle=0, radius=\diskradius cm];
        \drawtransformedcontour{1.00}{34.286}{0.00}
        \drawshadow{0.666667}{0}{0.0}
        \drawtransformedcontour{0.666667}{17.143}{0.6}
        \drawshadow{0.333333}{0}{0.6}
        \drawtransformedcontour{0.333333}{0}{1.0}

        \node[draw=none, anchor=center, inner sep=0pt]
            at (0,-1.20cm) {%
                \tikz[baseline=-0.5ex]{%
                    \node[draw=myred, circle, fill=myredlighter,
                          inner sep=1.0pt] {};%
                }%
                \hspace{0.5mm}\scalebox{0.7}{%
                    $=\textrm{pivot}(\w_i,\textrm{radius})$%
                }%
            };
    \end{scope}

    % Final samples after the radius-dependent chord transformations
    \begin{scope}[xshift=10cm]
        
        \draw[mygreylight, fill = mygreylighter]
            (-\diskradius cm,0)
            arc[start angle=180, end angle=0, radius=\diskradius cm];

        \begin{scope}[yscale=0.45]
            \fill[mygreylighter]
                (0,0) circle[radius=\diskradius cm];
            \foreach \rho in {0.333333,0.666667,1.00} {
                \draw[mygreylight]
                    (0,0) circle[radius={\rho*\diskradius cm}];
            }

            \drawfinalsamples{0.333333}{0}
            \drawfinalsamples{0.666667}{17.143}
            \drawfinalsamples{1.00}{34.286}
        \end{scope}

        \node[draw=none, anchor=center, inner sep=0pt]
            at (0,-1.20cm) {%
                \tikz[baseline=-0.5ex]{%
                    \node[draw=myorange, circle, fill=myorangelight,
                          inner sep=0.675pt] {};%
                }%
                \hspace{0.5mm}\scalebox{0.7}{$\sim f_r$}%
            };

    \end{scope}

\end{tikzpicture}
    \caption{ \label{fig:sampling} Direct BRDF sampling using radius-dependent azimuthal warps of the orthographic disk. }
\end{figure}
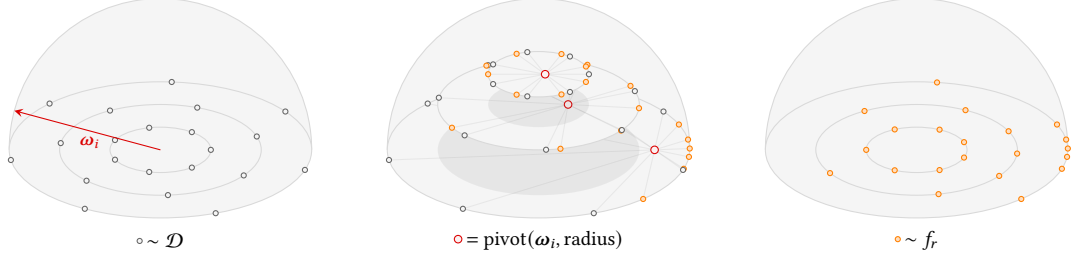

% =============================================================================================
\paragraph*{BRDF Sampling}
We leverage the McCullagh parameterization to sample the BRDF as illustrated in Figure~\ref{fig:sampling}.
The transformed point on the orthographic disk is
\begin{equation}
    \zeta_o
    :=
    x_o+\mathrm i y_o
    =
    r\frac{\eta-c}{1-\overline\eta\,c}.
    \label{eq:brdf-disk-sample}
\end{equation}
Here, the disk radius, the uniform point on its azimuthal isocontour, and the outgoing elevation are
\begin{equation}
    r
    =
    \sqrt{\xi_1},
    \qquad
    c
    =
    e^{2\pi\mathrm i\xi_2},
    \qquad
    z_o
    =
    \sqrt{1-r^2},
    \label{eq:uniform-disk-sample}
\end{equation}
where $\xi_1,\xi_2\in[0,1)$ are independent uniform random variables.
The Poisson parameter is constructed from the incident direction and the sampled radius as
\begin{equation}
    a_i
    :=
    \frac{x_i+\mathrm i y_i}{1+z_i}
    =
    q(z_i)e^{\mathrm i\phi_i},
    \qquad
    \eta
    :=
    -q(z_o)a_i
    =
    -\frac{r}{1+z_o}a_i.
    \label{eq:brdf-chord-pivot}
\end{equation}
Its magnitude and argument are $q(z_i)q(z_o)$ and $\phi_i+\pi$, respectively, as required by Equation~\eqref{eq:render-compact}.
Equation~\eqref{eq:brdf-disk-sample} preserves $r$ and warps only the associated azimuthal isocontour.
The angular Jacobian is exactly the BRDF density with respect to disk area:
\begin{equation}
    \frac{1}{\pi}
    \left|\frac{\partial\theta}{\partial\phi_o}\right|
    =
    2\Ppois\!\left(\frac{\zeta_o}{r};\eta\right)
    =
    f_r(\w_i,\w_o).
    \label{eq:brdf-jacobian}
\end{equation}
Finally, we lift $\zeta_o$ to the hemisphere:
\begin{equation}
    \w_o
    =
    \begin{pmatrix}
        \Re(\zeta_o)\\
        \Im(\zeta_o)\\
        z_o
    \end{pmatrix}.
    \label{eq:brdf-direction-sample}
\end{equation}
The resulting direction follows the target density
\begin{equation}
    f_r(\w_i,\w_o) \, z_o
    =
    \mathcal R(-\w_i,\w_o).
    \label{eq:brdf-sampling-density}
\end{equation}
Algorithm~\ref{alg:brdf-sampling} provides a possible implementation.

\begin{algorithm}[H]
    \begin{algorithmic}[1]
        \caption{Direct BRDF sampling}
        \label{alg:brdf-sampling}
        \Function{SampleBRDF}{$\xi_1$, $\xi_2$, $\w_i$}
        \State $(x_i,y_i,z_i)$ $\gets$ $\w_i$
        \State $r$ $\gets$ $\sqrt{\xi_1}$
        \State $z_o$ $\gets$ $\sqrt{1-r^2}$
        \State $c$ $\gets$ $e^{2\pi\mathrm i\xi_2}$
        \State $a_i$ $\gets$ $(x_i+\mathrm i y_i)/(1+z_i)$
        \State $\eta$ $\gets$ $-r a_i/(1+z_o)$
        \State $\zeta_o$ $\gets$ $r(\eta-c)/(1-\overline\eta\,c)$
        \State \Return $(\Re(\zeta_o),\Im(\zeta_o),z_o)$
        \EndFunction
    \end{algorithmic}
\end{algorithm}

% =============================================================================================
% =============================================================================================
\subsection{Sampling the Phase Function}

% =============================================================================================
\paragraph*{Elevation Sampling}
For the sake of completeness, we also use the same construction to sample the quadratic phase function.
Only the radial map and the resulting Poisson pivot need to be changed.
Let $\w_1=(x_1,y_1,z_1)$ be the current direction and let $\xi_1,\xi_2\in[0,1)$ be two independent uniform random variables.
We sample
\begin{equation}
    r=\sqrt{\xi_1},
    \qquad
    z_2=z_1+(1-z_1)r,
    \qquad
    r_2=\sqrt{1-z_2^2}.
    \label{eq:phase-sampling-elevation}
\end{equation}
Since $r$ has density $2r$, this radial map produces the elevation density
\begin{equation}
    2r
    \left|\frac{\partial r}{\partial z_2}\right|
    =
    \frac{2(z_2-z_1)}{(1-z_1)^2}
    =
    h(z_2\mid z_1).
    \label{eq:phase-radial-jacobian}
\end{equation}

% =============================================================================================
\paragraph*{Azimuthal Sampling}
For the sampled elevation, the transformed point on the corresponding azimuthal isocontour is
\begin{equation}
    \zeta_2
    :=
    x_2+\mathrm i y_2
    =
    r_2\frac{\eta-c}{1-\overline\eta\,c}.
    \label{eq:phase-sampling-direction}
\end{equation}
Here, the uniform point on the unit circle and the Poisson parameter are
\begin{equation}
    c
    =
    e^{2\pi\mathrm i\xi_2},
    \qquad
    \eta
    :=
    \frac{q(z_2)}{q(z_1)}
    \frac{x_1+\mathrm i y_1}{\sqrt{1-z_1^2}}
    =
    \frac{q(z_2)}{q(z_1)}e^{\mathrm i\phi_1}.
    \label{eq:phase-chord-pivot}
\end{equation}
Finally, we lift $\zeta_2$ to the sphere:
\begin{equation}
    \w_2
    =
    \begin{pmatrix}
        \Re(\zeta_2)\\
        \Im(\zeta_2)\\
        z_2
    \end{pmatrix}.
    \label{eq:phase-sampling-lift}
\end{equation}
The full Jacobian is therefore exactly the phase-function density
\begin{equation}
    h(z_2\mid z_1)
    \Ppois\!\left(\frac{\zeta_2}{r_2};\eta\right)
    =
    \Psi(\w_1,\w_2).
    \label{eq:phase-sampling-jacobian}
\end{equation}

% =============================================================================================
% =============================================================================================
% =============================================================================================
\section{Validation and Results}
\label{sec:results}

\begin{figure}
    \centering
    \includegraphics[width=\linewidth]{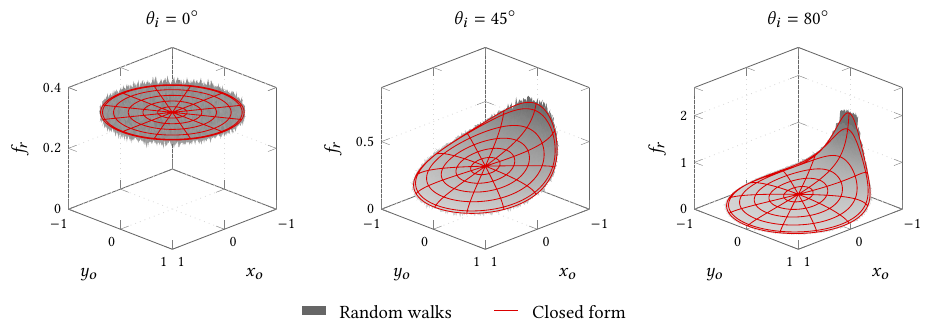}
    \caption{ \label{fig:numerical-tests} Random-walk histograms and the closed-form BRDF for three incident angles. }
\end{figure}

% =============================================================================================
\paragraph*{Numerical Validation}
We validate Equation~\eqref{eq:main-brdf} against random-walk histograms obtained with Algorithm~\ref{alg_walk} using the quadratic NDF.
Figure~\ref{fig:numerical-tests} compares these histograms with the closed-form expression for incident angles $\theta_i=0^\circ$, $45^\circ$, and $80^\circ$.
The expression agrees closely with the simulated data in all three configurations. Note that at normal incidence, the BRDF behaves exactly like a Lambertian 
BRDF, i.e., a constant. As incidence becomes increasingly grazing, the BRDF becomes more asymmetric and reflects more energy towards the 
opposite side of the disk. 

\begin{figure}
    \input{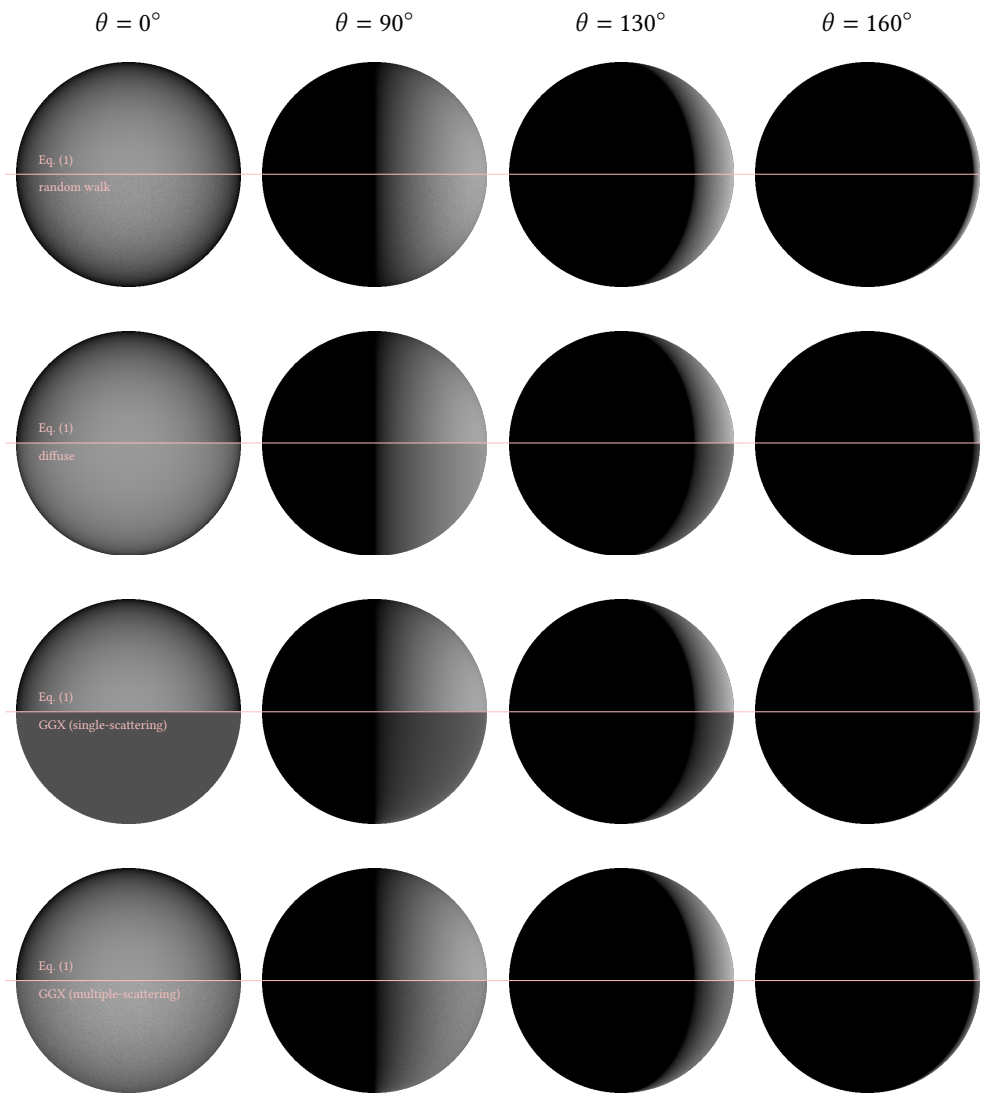}
    \caption{ \label{fig:directional-comparison} BRDF comparisons under directional illumination. }
\end{figure}

% =============================================================================================
\paragraph*{Directional Illumination}
We compare our BRDF with Lambertian reflectance and GGX microfacet BRDFs, in both single- and multiple-scattering variants.
The GGX models use unit roughness, $\alpha=1$, and all microfacets are perfectly reflecting. The single scattering model 
evaluates Equation~\eqref{eq:ggx1-f1}. 
Figure~\ref{fig:directional-comparison} shows these comparisons as a directional light rotates around the sphere.
Note that we also provide a comparison between Equation~\eqref{eq:main-brdf} and a random walk using 
Algorithm~\ref{alg_walk} to further validate our results. As can be seen, both renderings are in close agreement, with 
the difference residing in the variance of the random-walk approach. 

\begin{figure}
    \input{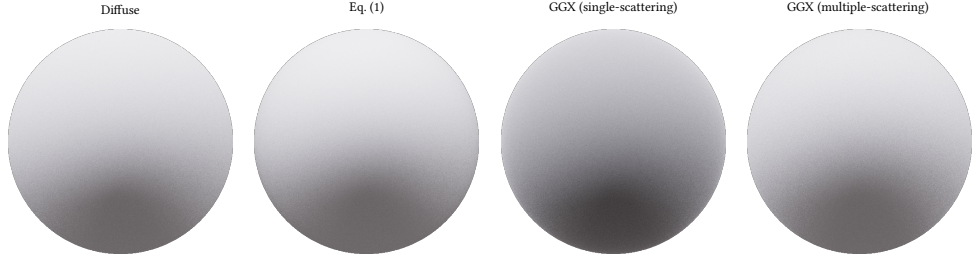}
    \caption{ \label{fig:sky-comparison} BRDF comparisons under image-based lighting using \texttt{uffizi.hdr}.
    Light probe courtesy of Paul Debevec, distributed by USC ICT at \protect\url{https://vgl.ict.usc.edu/Data/HighResProbes/}. }
\end{figure}

% =============================================================================================
\paragraph*{Image-Based Lighting}
Figure~\ref{fig:sky-comparison} compares the same four BRDFs under image-based lighting.
Under this illumination, our BRDF produces a diffuse-like appearance close to the Lambertian reference, while retaining a direction-dependent response.
The single-scattering GGX model appears darker because it neglects inter-reflections, whereas its multiple-scattering counterpart recovers the energy carried by these paths.

% =============================================================================================
% =============================================================================================
% =============================================================================================
\section{Conclusion}
\label{sec:conclusion}

% =============================================================================================
\paragraph*{Contributions}
We have introduced a one-sided quadratic NDF for which the complete random-walk response of a homogeneous semi-infinite microflake medium admits an elementary closed form.
Analytically marginalizing the intermediate directions and summing all scattering orders yields the energy-preserving BRDF of Equation~\eqref{eq:main-brdf}.
Its connection with the Poisson kernel provides direct importance sampling of both the BRDF and the phase function.
We validated our BRDF via random-walk histograms and rendering comparisons.

% =============================================================================================
\paragraph*{Limitation and Outlook}
These results settle the existence of an NDF capable of producing an elementary closed-form response for random walks in homogeneous microflake media.
However, the proposed NDF is limited in the range of appearances it can represent as it lacks the typical roughness parameters of a microfacet NDF.
It thus remains to be determined whether a parametric model can preserve such a kind of closed form while offering a parameter that plays the 
role of roughness.

% =============================================================================================
% =============================================================================================
% =============================================================================================
\section*{Acknowledgements and Backstory}
This paper concludes a project I started in March 2022.
I thank Eugene d'Eon for encouraging me to pursue this work at that time when I had most of the derivations 
in place, except for the quadratic NDF.
Back then, I believed that if an elementary closed form could not be obtained from the simplest 
uniform one-sided NDF, there was no hope for any other NDF.
I failed to find such a closed form and stopped working on the project.
On the positive side, I realised that the uniform NDF was the GGX at unit roughness, 
and that led me to write an importance sampling algorithm for the GGX VNDF.
A few years later came ChatGPT 5.6, and I decided to challenge it using my notes. It proved my original 
assumption wrong: after its own 
unsuccessful attempts with GGX, ChatGPT 5.6 found the quadratic NDF when I asked it to 
find any NDF that leads to a closed-form BRDF. Credit to OpenAI for finding this NDF: your product 
is changing the way knowledge is made.
I thank my wife for letting me write the (very many!) prompts that led to this finding during my parental 
leave and dedicate the article to my children as it settles a question that was important to me.

% =============================================================================================
% =============================================================================================
% =============================================================================================
\bibliography{multiscattering}{}
\bibliographystyle{plain}
\end{document}